\documentclass[iop]{emulateapj}   
\usepackage{pdflscape}
\usepackage{graphicx}

\begin{document}

\def\chandra {{\it Chandra}~}
\def\xmm     {{\it XMM-Newton}~}
\def\etal   {{et~al.}~}
\def\msun   {{$\rm{M}_{\odot}$}}
\def\wav {{\sc WAVDETECT}~}
\newcommand{\nh}{$N_{\rm H}$}
\newcommand{\sig}{$\sigma$}
\newcommand{\mbh}{$M_{\rm BH}$}
\newcommand{\es}{erg s$^{-1}$}
\newcommand{\esc}{erg s$^{-1}$ cm$^{-2}$}
\newcommand{\Msun}{M$_{\odot}$}
\newcommand{\chired}{$\rm{\chi^2_{red}}$}
\newcommand{\wave}{{\sc Wavdetect}}

\title{Discovery of a Large Population of Ultraluminous X-ray Sources in the Bulge-less Galaxies NGC 337 and ESO 501-23}

\author{Garrett Somers\altaffilmark{1},
 Smita Mathur\altaffilmark{1,2},
 Paul Martini\altaffilmark{1,2},
 Linda Watson\altaffilmark{3},
 Catherine J. Grier\altaffilmark{1} and
 Laura Ferrarese\altaffilmark{4}}

\altaffiltext{1}{Department of Astronomy, The Ohio State University, 140 W. 18th Ave, Columbus, OH 43210; somers@astronomy.ohio-state.edu}
\altaffiltext{2}{Center for Cosmology and Astroparticle Physics, The Ohio State University, 191 W Woodruff Avenue, Columbus, OH 43210, USA}
\altaffiltext{3}{Harvard-Smithsonian Center for Astrophysics, 60 Garden Street, Cambridge, MA 02138, USA}
\altaffiltext{4}{Hertzberg Institute of Astrophysics, 5071 West Saanich Road, Victoria, BC, V9E 2E7, Canada}

\begin{abstract}

We have used \chandra observations of eight bulge-less disk galaxies to identify new ultraluminous X-ray source (ULX) candidates, study their high mass X-ray binary (HMXB) population, and search for low-luminosity active galactic nuclei (AGN). We report the discovery of 16 new ULX candidates in our sample of galaxies. Eight of these are found in the star forming galaxy NGC 337, none of which are expected to be background contaminants. The HMXB luminosity function of NGC 337 implies a star formation rate (SFR) of 6.8$^{+4.4}_{-3.5}$ \msun\ yr$^{-1}$, consistent at 1.5$\sigma$ with a recent state of the art SFR determination. We also report the discovery of a bright ULX candidate (X-1) in ESO 501-23. X-1's spectrum is well fit by an absorbed power law with $\Gamma = 1.18^{+0.19}_{-0.11}$ and N$\rm{_H}$ = 1.13$^{+7.07}_{-1.13} \times 10^{20}$ cm$^{-2}$, implying a 0.3-8 keV flux of $1.08^{+0.05}_{-0.07} \times 10^{-12}$ \esc. Its X-ray luminosity (L$_X$) is poorly constrained due to uncertainties in the host galaxy's distance, but we argue that its spectrum implies L$_X > 10^{40}$ \es. An optical counterpart to this object may be present in HST imaging. We also identify ULX candidates in IC 1291, PGC 3853, NGC 5964 and NGC 2805. We find no evidence of nuclear activity in the galaxies in our sample, placing a flux upper limit of $4 \times 10^{-15}$ \esc\ on putative AGN. Additionally, the type II-P supernova SN 2011DQ in NGC 337, which exploded 2 months before our X-ray observation, is undetected.

\end{abstract}

\section{Introduction}

Ultraluminous X-ray sources (ULXs) are extra-Galactic, off-nuclear X-ray objects which radiate at $\gtrsim 2 \times 10^{39}$ \es\ in the 0.3-10 keV range, assuming isotropic emission (for a recent review, see Feng \& Soria 2011). This is a remarkable luminosity, as it represents the Eddington limit for an accretion disk around the most massive stellar mass black holes (StMBHs) known in the Galaxy ($\sim$ 20 \msun; e.g. Remillard \& McClintock 2006). Over one hundred ULX candidates have now been identified in the local universe (Winter \etal 2006, Swartz \etal 2011), though the number with well-sampled spectra remains low. The most natural explanation for these objects are accreting black holes with masses in the range $10^2-10^5$ \msun, often referred to as intermediate mass black holes (IMBHs; Colbert \& Mushotzky 1999). This explanation gained popularity because it does not violate the Eddington limit, and explains ULXs in the well known framework of black hole binaries. Indeed, early investigations of ULXs found evidence of cool accretion disks and, in some cases, quasi-periodic oscillations (QPOs) in their spectra, both indicative of IMBH binaries (see Miller \& Colbert 2004 for a review). In particular, the object ESO 243-49 HLX-1 is one of the strongest known IMBH candidates (Farrell \etal 2009). However, recent arguments, such as the unphysically large amount of mass required to reproduce the ULX populations in some galaxies (King \etal 2004, see also Roberts 2007), have disfavored the theory that IMBHs dominate the ULX population. Other authors have suggested strong beaming from sub-Eddington accretion onto a StMBH (e.g. King \etal 2001), although the discovery of emission nebulae around several ULXs argues for mostly isotropic emission (Pakull \& Mirioni 2003).

An alternative explanation for ULXs involves super-Eddington accretion onto a stellar mass black hole, perhaps with mild beaming (e.g. King \etal 2001, Begelman 2002). This requires a different accretion paradigm than the classic thin disk models (Shakura \& Sunyaev 1973), such as the class of optically thick, slim disk models (e.g. Abramowicz \etal 1988, Watari 2000, Poutanen \etal 2007), which can radiate at several times Eddington. This may result in a previously unknown black hole binary (BHB) accretion state dubbed the ultraluminous state (Gladstone \etal 2009) which may characterize StMBH binaries for a fraction of their lifetime. In this frame work, ULXs would represent the high-mass end of the high mass X-ray binary (HMXB) distribution, radiating in their most extreme state. This idea is supported by the discovery of luminous blue stellar counterparts to several ULXs, a necessary component of a HMXB (Liu \etal 2004, Levan \& Goad 2008, Jonker \etal 2012). Furthermore, ULXs are preferentially found in regions of recent star formation and low metallicity (Kaaret 2005, Swartz \etal 2009, Pakull \& Mirioni 2002), likely nurseries of massive StMBHs and binary companions (Belczynski \etal 2010). Only dynamical mass measurements can definitively confirm the nature of these mysterious objects, so for now the controversy persists.  

If some ULXs contain IMBHs, their study may provide insight into the formation of super massive black holes (SMBH), which likely grow from intermediate mass seeds. The relationship between IMBHs and SMBHs is of great importance to the study of galaxy evolution, as observations of active galactic nuclei (AGN) and their host environments provide strong evidence that galaxies and their central black holes co-evolve. In particular, the \mbh--$\sigma \rm{_{Bulge}}$ relationship and the \mbh--$M\rm{_{Bulge}}$ relationship (Ferrarese \& Merritt 2000, Gebhardt \etal 2000, H\"{a}ring \& Rix 2004) imply that the evolutionary history of a galaxy and its central massive black hole may be connected. Since these correlations are fundamental to our understanding of hierarchical galaxy formation and evolution in the $\Lambda$CDM universe (e.g. Menci \etal 2004), the full parameter space of \mbh, $M \rm{_{Bulge}}$, and $\sigma \rm{_{Bulge}}$ must be explored. Although the \mbh--$\sigma \rm{_{Bulge}}$ relation holds in many mass regimes, it should break down in bulge-less galaxies if the formation of a SMBH is connected to the presence of a bulge. Yet in some cases, SMBHs have been identified in bulge-less galaxies by their X-ray AGN signatures (Filippenko \& Ho 2003, Satyapal \etal 2007, Shields \etal 2008, Ghosh 2009, Secrest \etal 2012, Araya Salvo \etal 2012). This has led some to hypothesize that SMBHs in late type galaxies are formed through different processes than black holes associated with spheroidal systems (Kormendy \& Bender 2011). Bulge-less galaxies therefore provide a crucial laboratory for studying galaxy growth in a regime where our understanding of SMBH/galaxy co-evolution may differ from the standard model (Mathur \etal 2012). Although statistics about the prevalence of bulge-less AGN are beginning to accumulate, more late type galaxies must be observed to probe the low mass end of the \mbh--$\sigma \rm{_{Bulge}}$ relation.

In this paper we discuss \chandra observations of eight nearby, late-type galaxies (see Table 1) selected from among the twenty objects studied in Watson \etal (2011, W11 hereafter). W11 conducted a survey in a variety of wavelengths to study the effects of rotation speed on star formation in bulge-less galaxies (see also Watson \etal 2012). In particular, their sample was selected to straddle the critical 120 km s$^{-1}$ rotation speed where the properties of the cold ISM appear to change abruptly (Dalcanton \etal 2004). From their set of twenty, we selected eight galaxies that span the full range of mass, star formation rate, and rotational velocity explored in their study. We report here our findings from the analysis of these observations. In $\S$2, we briefly describe the \chandra observations and our reduction and analysis methods. $\S$3 contains a description of the X-ray populations within each galaxy in the sample, and highlights potentially interesting sources. $\S$4 presents the results of our search for nuclear activity, and describes the ULX candidate populations of NGC 337 and ESO 501-23. Finally, $\S$5 summarizes our conclusions. Throughout this paper, we make use of the log$N-$log$S$ formulation of Moretti, Campana, Lazzati and Tagliaferri (2003, M03 hereafter)\begin{footnote}
{
For the soft band 0.5-2.0 keV, and the hard band 2.0-10.0 keV, the number of sources expected, per square degree, above flux limit S (cgs):
N$_S$($>$S) = $\rm{\frac{1.09 \times 10^{-23}}{S^{1.82}\ +\ 1.34 \times 10^{-17}\ S^{0.60}}}$ and
N$_H$($>$S) = $\rm{\frac{4.43 \times 10^{-20}}{S^{1.57}\ +\ 6.14 \times 10^{-17}\ S^{0.44}}}$.
}\end{footnote}
as a rough estimate of the expected number of cosmic X-ray background (CXB) objects above a given flux limit.

\section{Observations \& Data Reduction}

The observations were undertaken with the \chandra X-ray Observatory (Weisskopf \etal 2000) Advanced CCD Imaging Spectrometer (ACIS) between November 2010 and January 2012. Each galaxy was imaged for 9.93 ks using the ACIS S-array chips S2 through S4, with the galaxy centered on the back illuminated S3 chip. Following the observations, \chandra X-ray FITS files were downloaded from the \emph{Chandra Data Archive} and analyzed with the \chandra Interactive Analysis of Observations suite (CIAO; Fruscione \etal 2006) version 3.4. Following the method of Ghosh \etal (2008), we produced level 2 event files from the acquired level 1 event files by applying grade, status, and good time interval filters, and limiting our observations to the 0.3-8 keV range. We searched for contaminating background flares by binning the exposure of each chip into 30 time bins ($\sim$ 300s each). If the count rate of a bin deviated from the mean by more than 10\%, that bin was excluded from further analysis. The details of each observation, including the background flare-subtracted exposure times, are listed in Table 2. 

Next, we used the CIAO task \wave\ on each chip to identify potential X-ray objects. At the location of each candidate, counts were extracted from a 95\% energy enclosed circle, 2.3$''$ for on-axis point sources, and larger for off-axis point sources. The background contribution was removed using an annulus with an inner (outer) radius of 2$\times$ (5$\times$) the source extraction circle's radius. For sources that lay on the edge of the CCD, a half annulus was used to determine the background and the effects of off-CCD dithering were taken into account. If another detected source fell within the annulus, it was excised before the count rate was determined. \wave\ sources with less than 5 counts, or with less than a 2\sig\ deviation above the background flux, were rejected. These criteria were chosen to ensure minimal false positives, which can be numerous in low signal-to-noise data. Absolute astrometric errors are less than 0.8$''$ for 99\% of \chandra sources. We tested this by examining the relative astrometry of sources with obvious SDSS counterparts, and found that typical offsets were within this error.

\begin{figure*}
\begin{center}
{\setlength{\fboxsep}{0mm}
 \setlength{\fboxrule}{2pt}
\fbox{\includegraphics[width=2in]{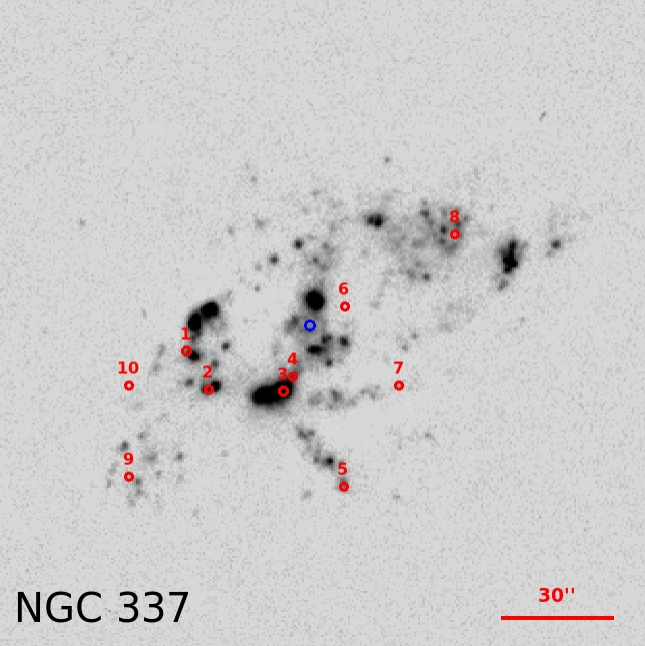}}}
{\setlength{\fboxsep}{0mm}
 \setlength{\fboxrule}{2pt}
\fbox{\includegraphics[width=2in]{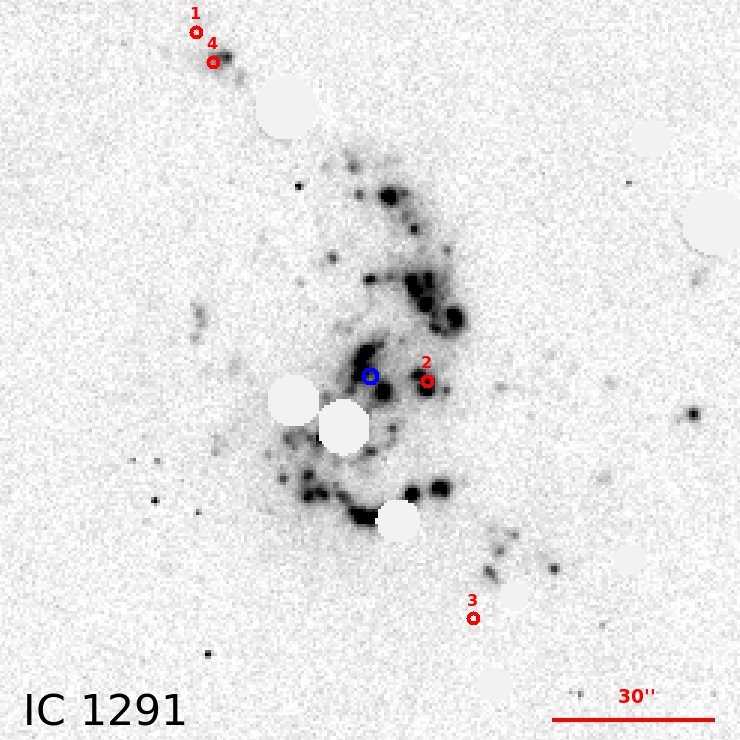}}}
{\setlength{\fboxsep}{0mm}
 \setlength{\fboxrule}{2pt}
\fbox{\includegraphics[width=2in]{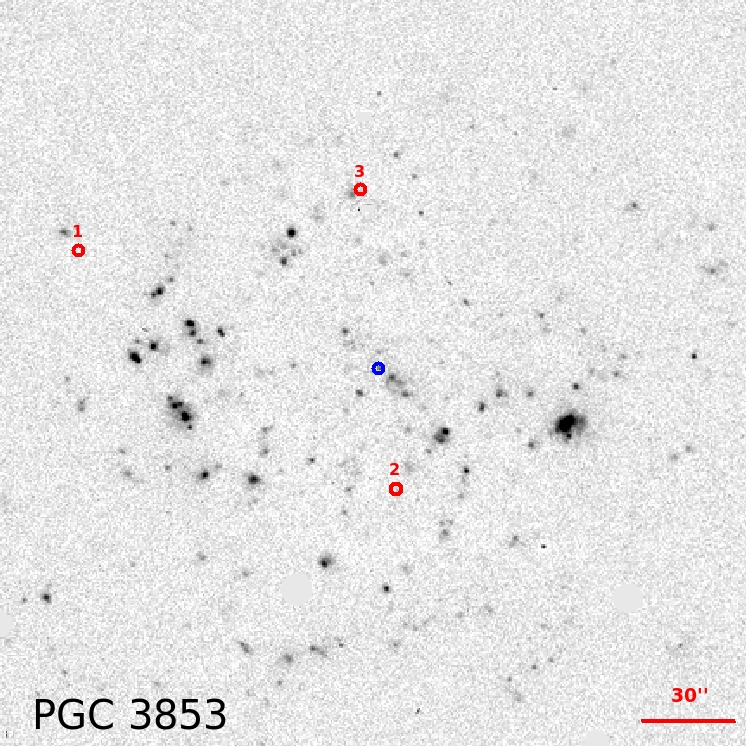}}}

\vspace{2mm}
{\setlength{\fboxsep}{0mm}
 \setlength{\fboxrule}{2pt}
\fbox{\includegraphics[width=2in]{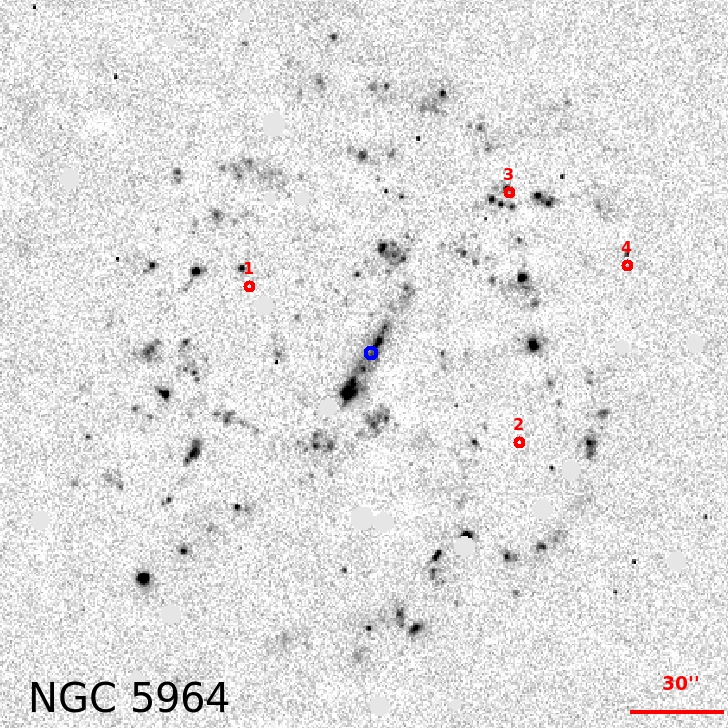}}}
{\setlength{\fboxsep}{0mm}
 \setlength{\fboxrule}{2pt}
\fbox{\includegraphics[width=2in]{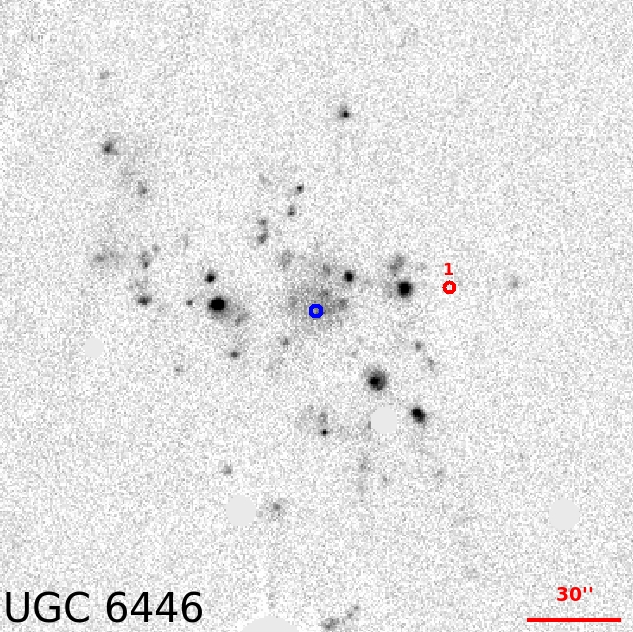}}}
{\setlength{\fboxsep}{0mm}
 \setlength{\fboxrule}{2pt}
\fbox{\includegraphics[width=2in]{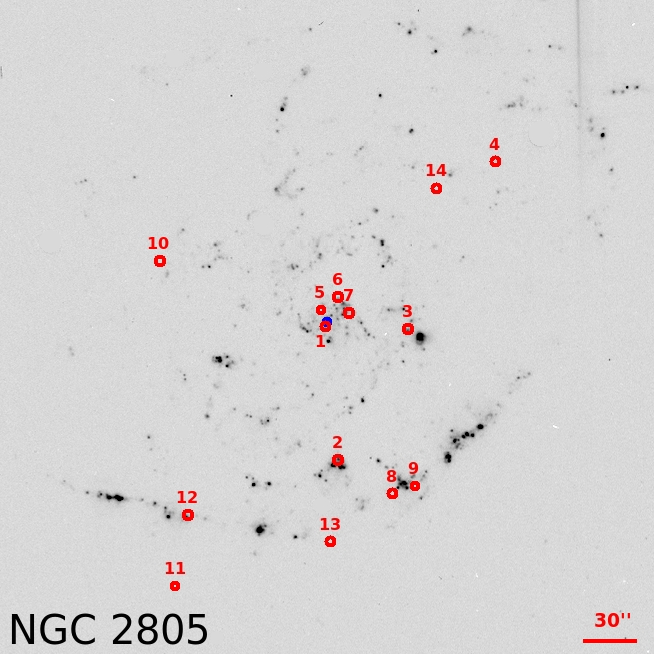}}}

\vspace{2mm}
{\setlength{\fboxsep}{0mm}
 \setlength{\fboxrule}{2pt}
\fbox{\includegraphics[width=2in]{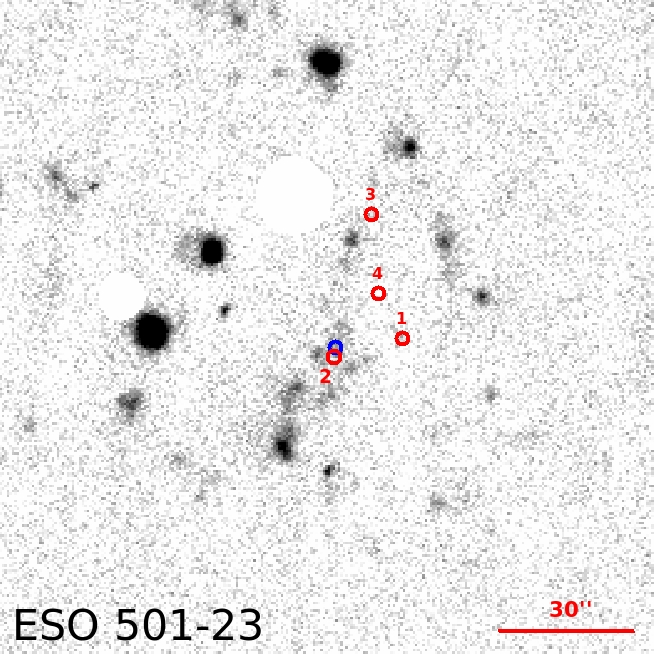}}}
{\setlength{\fboxsep}{0mm}
 \setlength{\fboxrule}{2pt}
\fbox{\includegraphics[width=2in]{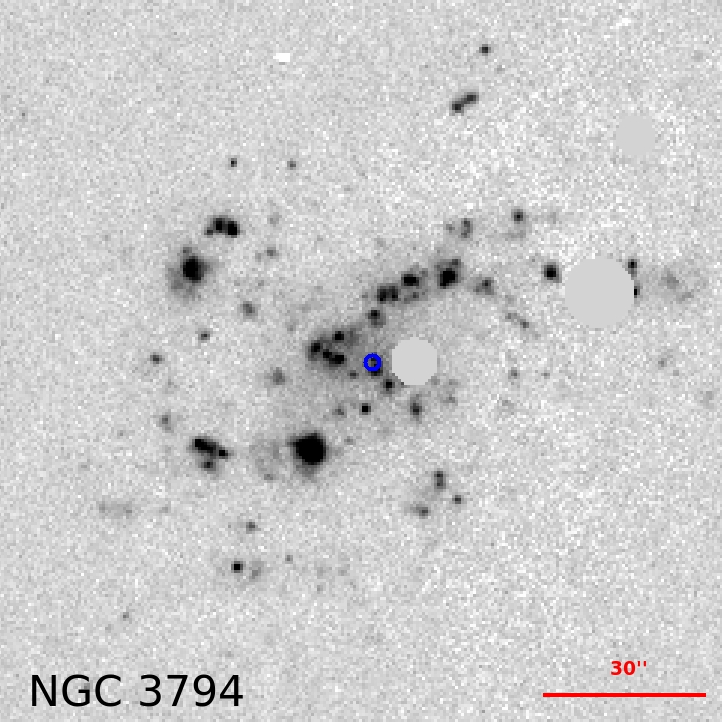}}}

\caption{X-ray source locations (red points) from Table 7, superimposed on H$\alpha$ images (gray scale) from the 2.4m Hiltner Telescope at MDM Observatory (see W11 for details). Blue circles represent the central coordinates listed in Table 1. }
\end{center}
\end{figure*}

\section{X-Ray Results}

\wave\ identified a total of 190 candidate sources within our eight fields. Sixteen of these were discarded based on our rejection criteria, leaving the 174 objects listed in Table 3. These objects range in nature from background galaxies, to foreground stars, to X-ray binaries within the targeted galaxies. To identify potential counterparts to these sources, we cross correlated our sample with the USNO-B1.0, SDSS DR8, and WISE point source catalogs using the \emph{Vizier}\footnote{http://vizier.u-strasbg.fr/viz-bin/VizieR} database (Monet \etal 2003, Adelman-McCarthy \etal 2011, Wright \etal 2010). Objects falling within 2$''$ of a source in Table 4 were included. \emph{Chandra}'s 90\% source location on-axis error circle is 0.6$''$, and 99\% of all sources fall within 1$''$, so more distant objects are unlikely to be physically related. Nevertheless, we include them in our cross-correlation table to account for the larger positional error of off-axis sources, and the possibility of significant proper motion between epochs. The resulting source lists are in Tables 4, 5, and 6.

Forty X-ray sources are detected within the combined 25th mag arcsec$^{-2}$ B-band isophotes (D25) of the eight galaxies in the sample (see Table 1), including 18 with fluxes consistent with ULXs. These sources are listed in Table 7. H$\alpha$ images of the galaxies are shown in Figure 1, over plotted with X-ray source locations. Due to the high probability of CXB contamination, each source must be evaluated individually to determine its nature. This is possible in some cases when the object is sufficiently bright, or when counterparts can be identified. We briefly describe the findings for each galaxy here, and highlight potentially interesting sources.

\subsection{NGC 337}
NGC 337 is an SBd galaxy at a distance of 20.7 Mpc (see Table 2). This object is part of the SINGS sample of galaxies (Kennicutt \etal 2003), and  has been extensively studied in the literature. Our exposure uncovered ten sources within the D25 ellipse, including eight whose fluxes are consistent with ULXs, as seen in Figure 1. M03 predict $\sim$ 0.6 soft and $\sim$ 1.3 hard background sources above our flux limit within this area, which implies the existence of a large population of ULXs associated with NGC 337. These objects will be examined further in $\S$4.2. Additionally, this \chandra observation was taken $\sim$ 2 months after the discovery of the type II-P supernova SN 2011DQ in NGC 337. This event was located at R.A. = $\rm{00^h59^m47.75^s}$, Dec. = $\rm{07^d34^m20.5^s}$ (J2000), and is clearly within the galactic disk (Barbon \etal 2009). No counts above the background rate were detected at this location, corresponding to a 3\sig\ upper limit of 2.8$\times$10$^{38}$ \es\ from 0.3-8 keV. This is not surprising: although several type II-P supernovae have been detected in X-rays, such as SN 2004dj (Pooley and Lewin 2004), their X-ray luminosities at a distance of 20.7 Mpc would be below the detection limit at the time of our observation.

\subsection{IC 1291}
IC 1291 is an SBdm galaxy at a distance of 31.5 Mpc. We find four sources in this galaxy, two of which (X-1 and X-2) have fluxes consistent with ULXs at high significance ($>3\sigma$), and two (X-3 and X-4) with moderate significance ($\sim$ 1$\sigma$). X-1 and X-4 lie outside the D25 diameter listed in RC3, but within the D25 diameter listed in other catalogs (e.g. Tully 1988), so we include them in our table. H$\alpha$ observations imply that this galaxy is in an epoch of star formation, and thus is expected to host a few bright X-ray sources. X-2 is perhaps the best ULX candidate, as it coincides with a bright H$\alpha$ region and has a bright optical counterpart in HST images. However, due to the considerable distance of IC 1291 and the low signal-to-noise of our observation, only 28 net counts were observed for this object. Deeper exposures are therefore needed to confirm its nature. We caution that the two dimmer sources, despite being at the ULX flux level, are also consistent with CXB contaminants. This is because only 8.5 net counts are are needed to infer a luminosity in excess of 10$^{39}$ \es\ at 31.5 Mpc, and M03 predict $\sim$ 1 CXB source in this observation.

\subsection{PGC 3853}
PGC 3853 is an SABd galaxy at a distance of 11.4 Mpc. We find three sources in this galaxy, two that are dim ($\sim$ 5 cts) and one that is bright ($\gtrsim$ 150 cts). M03 predict $\sim$ five CXB sources in this observation, implying that some or all PGC 3853 detections may be contaminants. Using H$\alpha$ and HST images, we find no obvious optical counterparts to the dimmer sources, X-2 and X-3. Considering their low fluxes, we conclude they are likely distant background galaxies. X-1 is bright enough that it is likely associated with PGC 3853, but it does not lie in an area of recent star formation, and does not have any obvious optical counterparts in HST imaging or our cross-correlation tables. It is still possible that this source is a ULX, as Swartz \etal (2009) found that only $\sim$ 60\% of ULXs are associated with areas of recent star formation.

\subsection{NGC 5964}
NGC 5964 is an SBd galaxy at a distance of 24.7 Mpc. There are four sources within the D25 ellipse of this galaxy, two (X-3 and X-4) that are below the ULX threshold ($\lesssim$ 10 cts), and two (X-1 and X-2) that are above ($\gtrsim$ 20 cts). M03 predict $\sim$ 5 CXB sources in this observation, and patchy H$\alpha$ emission implies an absence of large scale star formation in the recent past, so these sources are consistent with background contaminants. However, X-3 is relatively close to a region with a young stellar population, and might be associated with a black hole binary that has now traveled away from its birth location. The projected separation between the X-ray source and the nearby H$\alpha$ source is $\sim$ 150 pc, and could be covered within a few 10s of Myrs at reasonable speeds. The two brightest sources, NGC 5964 X-1 and X-2, have no obvious visible counterparts or H$\alpha$ association, so their nature remains unclear.

\subsection{UGC 6446}
UGC 6446 is an SAd galaxy at a distance of 18.0 Mpc. Our exposure reveals one dim source within the D25 ellipse. We detect only 4.7 net counts, near our detection threshold but statistically significant. If this source is real, it could be a high-mass X-ray binary within UGC 6446 or, more likely, a background contaminant. As this galaxy is weak in H$\alpha$, the lack of notable X-ray sources is not surprising.

\subsection{NGC 2805}
NGC 2805 is an SABd galaxy at a distance of 28.0 Mpc. This galaxy has the largest population of X-ray point sources in our sample, with fourteen detections inside D25 and five consistent with ULXs. It also has the largest projected area in our sample, and M03 predict twelve background sources above our detection limit, including five consistent with ULXs. Due to the large expected CXB contribution, it is difficult to draw conclusions about the X-ray population of this galaxy. However, we note that ten of the fourteen sources appear consistent with areas of recent star formation, a fraction that would be difficult to produce with randomly oriented background sources. Additionally, four of the sources lie within the central 15$''$ of the galaxy, another unlikely orientation for background sources. Finally, the H$\alpha$ measurements of W11, the H$\alpha$/SFR relation of Kennicutt (1998), and the SFR/N($L > 2 \times 10^{38}$ \es) relation of Grimm, Gilfanov, and Sunyaev (2003, G03 hereafter) predict four X-ray sources above our flux limit within the D25 ellipse of NGC 2805. This implies that eleven of the sources are background, consistent with the prediction of M03. For these reasons, we conclude that several of these bright X-ray sources are likely associated with NGC 2805. X-3, X-4 and X-8 in particular are strongly consistent with ULXs, and are the best candidates in this observation.

\begin{figure}
\begin{centering}
{\setlength{\fboxrule}{1.5pt}
\fbox{\includegraphics[width=3in]{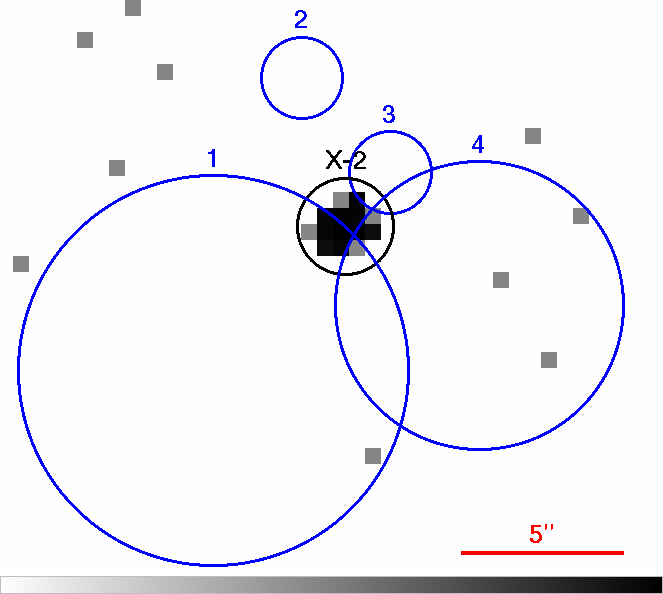}}
}
\caption{Unsmoothed image of ESO 501-23 X-2, shown with four reported central coordinates of the host galaxy. In order, \#1 is from RC3 (de Vaucouleurs \etal 1991), \#2 is from the HYPERLEDA catalog (Paturel \etal 2003), \#3 is from the 6 degree field survey (Jones \etal 2005), and \#4 is from W11. The first three are photometric centers, while the fourth is a dynamical center. The size of each circle represents the positional uncertainty of the determination. Up is North and left is East.}
\vspace{3mm}
\end{centering}
\end{figure}

\subsection{ESO 501-23}
ESO 501-23 is an SBdm galaxy at an adopted distance of 7.01 Mpc, although this distance is uncertain by a factor of two (see $\S$4.3). We find a total of four objects within this galaxy, including the brightest X-ray source discovered in this survey, ESO 501-23 X-1. This object is discussed further in $\S$4.3. The source X-2 lies only $\sim$ 2$''$ from the galactic center listed in Table 2. However, because of the peculiar shape of this galaxy, there is significant disagreement about the location of the galactic center, and several independent determinations of the central coordinates exist, as shown in Figure 2. Therefore, we do not claim this object is an AGN, although this possibility should not be discarded given its proximity to the central region. We can identify X-3 as a nearby star seen in HST imaging, and recorded in the Carlsberg-Meridian Catalog (CMC 2006). X-4 has no identifiable optical or IR counterpart, so it may be a HMXB or a CXB source.

\subsection{NGC 3794}
NGC 3794 is an SABd galaxy at a distance of 19.2 Mpc. No \chandra sources were identified within the D25 ellipse of this galaxy, moderately consistent with the M03 prediction of $\sim$ 1.3 sources above the flux limit.

\section{Discussion}

\subsection{AGN Candidates}

We searched the central regions of the eight galaxies for evidence of AGN activity. There is considerable difficulty in determining the central coordinates of bulge-less galaxies, which often lack an easily identifiable central structure. In particular, dynamical measurements of the central coordinates disagree with photometric measurements at the $\gtrsim$ 5$''$ level for NGC 2805 and ESO 501-23. Inner and outer photocenters may differ as well. This makes AGN detection claims difficult, as sub-arcsecond confidence is required to confirm nuclear activity. In Table 1 we adopt photometric centers from 2MASS, SDSS and 6dF as the galactic central coordinates. AGN at these coordinates, as well as the kinematic centers determined by W11, with $\mathit{f}\rm{_{0.3-8 keV}} \gtrsim\ 4 \times 10^{-15}$ \esc\ are ruled out with 2\sig\ confidence in all cases, with the possible exception of ESO 501-23 (see $\S$3.7). The corresponding luminosities for each galaxy are shown in column 11 of Table 1. In addition, we stacked the galactic nuclei to search for a cumulative detection, but none was found.

     \subsection{The ULXs in NGC 337}

There are 8 sources within the inner 1$'$ of NGC 337 that have fluxes comparable to ULXs at a distance of 20.7 Mpc (see Figure 1). These include all NGC 337 sources listed in Table 7, except for X-2 and X-10. Based on M03, we expect $\sim$ 0.35 background sources in the central arcminute above the ULX threshold. It is therefore very likely that all eight of these candidates are ULXs associated with NGC 337. In addition, 7 of the 8 objects coincide with H$\alpha$ emission, strengthening the ULX interpretation (Swartz \etal 2009). The $2\sigma$ detection limit of our \chandra image is $6 \times 10^{38}$ \es, lower than the ULX lower bound of $10^{39}$ \es. Therefore our sample represents the full population of ultraluminous objects present in the galaxy at the time the data were taken. However, the large variability of ULX luminosities, in some cases over orders of magnitude (Middleton \etal 2012), implies that future observations could reveal new ultraluminous X-ray sources.

We can use our observations to evaluate the star formation properties of NGC 337. G03 describe a correlation between a galaxy's star formation rate (SFR) and the luminosity function (LF) of its high mass X-ray binaries, defined in this case as X-ray binaries with a massive stellar companion, and an X-ray luminosity $>$ 2 $\times$ 10$^{38}$ \es. This can be explained as a result of X-ray binaries created during epochs of star formation that accrete their companions on a timescale similar to the lifetime of star formation indicators. They found that the LF was best fit by a cut-off power law scaled linearly with SFR:
\begin{equation}
\rm{\frac{d\emph{N}}{d\emph{L}_{38}} = (3.3^{+1.1}_{-0.8})}\ \rm{SFR}\ L^{-1.61 \pm 0.12}_{38}\ \ \ for\ \ \ L\ <\ L_C
\end{equation}
\begin{equation}
\rm{N(>L) = 5.4\ SFR\ (L_{38}^{-0.61} - 210^{-0.61})}
\end{equation} \\
Here, SFR is in units of \msun\ yr$^{-1}$, L$_{C}$ $\sim$ 2 $\times$ 10$^{40}$ \es, and L$\rm{_{38}}$ $\equiv$ $\rm{\frac{L}{10^{38}\ erg\ s^{-1}}}$. The upper bound of L$_C$ is a result of the observed power law break in the HMXB LF above $\sim$ 2 $\times$ 10$^{40}$ \es\ (e.g. Swartz \etal 2011). The substantial error bars in Eq. (1) encompass several effects. One such effect is the inhomogeneity of HMXB LFs used to derive this relation. G03 took ULX luminosities directly from the literature for the galaxies considered in their study. Each study had its own methods of determining the target galaxy's HMXB LF (see the references in Table 1 of G03 for more information), though most assumed an absorbed power law. Band passes ranged from 0.7-7 keV to 0.1-10 keV, and several different spectral indices were used. This leads to a scatter in inferred flux for a fixed count rate. Another effect is the treatment of extinction. In most cases, Galactic hydrogen was the only extincting medium accounted for. This ignores absorption internal to the target galaxy, which may contribute to net extinction and spectral hardening. To complicate matters further, this can vary significantly between objects within the same galaxy. Uncertainties in galactic distances and star formation rates also contribute, and may be the dominate sources of error. Considering these effects, G03 found an error in derived star formation rate of $\sim$ 50\%. Mineo, Gilfanov and Sunyaev (2012) recently repeated this analysis with closer attention to the systematic errors affecting G03, but found nearly the same uncertainty in derived SFR, implying this scatter may be real. With this precaution, we can proceed using the relation of G03.

\begin{figure*}
\centering
\includegraphics[width=6.5in]{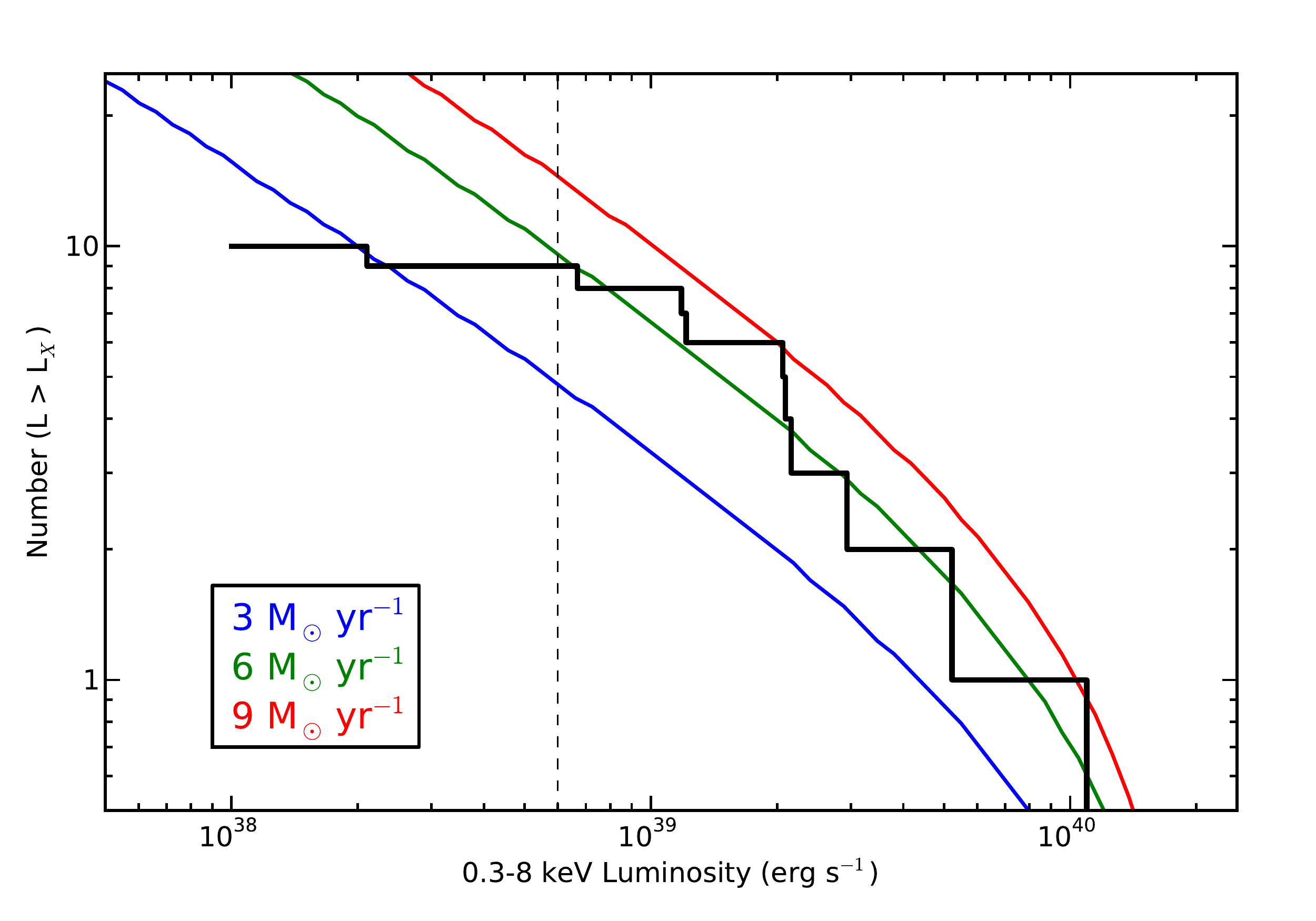}
\caption{X-ray luminosity function of NGC 337, in black. X-ray luminosity values are determined with a power law with galactic absorption, $\Gamma = 1.7$, and a band-pass of 0.3-8 keV. Over-plotted are 3 realizations of Eq. (2), for star formations rates of 3, 6, and 9 \msun\ yr$^{-1}$.}
\end{figure*}

To derive the HMXB LF, we choose a band-pass of 0.3-8 keV, a spectral index of 1.7, and Galactic absorption. This falls roughly in the middle of the literature range studied by G03. Only sources with L$_X > 6 \times 10^{38}$ \es\ are considered to ensure completeness, though we note this may contribute to our total error. Using a maximum likelihood fit, we determine a SFR of 6.8$^{+4.4}_{-3.5}$ \msun\ yr$^{-1}$. Figure 3 shows the HMXB luminosity function of NGC 337, along with realizations of Eq.(2) for three SFRs. To evaluate whether background contaminants could affect this value, we randomly subtracted single objects from the LF and re-derived the best fit. We found that the change in derived star formation rate was at most 1 \msun\ yr$^{-1}$. This is a small effect compared to the systematic uncertainties, and likely even smaller since the most probable contaminants are lower flux sources which do not affect the LF as dramatically. 

We can compare this estimate with SFRs derived from photometric indicators. G03 used the H$\alpha$ and far-IR SFR relations of Rosa-Gonz\'{a}lez \etal (2002), so we employ these for consistency. Watson \etal (2012) and Kennicutt \& Moustakas (2006) found an extinction-corrected H$\alpha$ flux of $1.86 \times 10^{-12}$ \esc, and $3.86 \times 10^{-12}$ \esc, giving SFR($\rm{H\alpha}$) $\sim$ 1.0 and 2.2 \msun\ yr$^{-1}$, respectively. The far-IR luminosity measurement of Sanders \etal (2003), rescaled to a distance of 20.7 Mpc, gives a luminosity of $\sim 4.1 \times 10^{43}$ \es. This implies SFR(FIR) $\sim$ 1.9 \msun\ yr$^{-1}$. Extinction in the FIR is minimal, so we do not correct our estimate. To determine if these values are consistent with a state-of-the-art calculation, we rescaled the recent analysis of \emph{Spitzer} and H$\alpha$ images by Calzetti \etal (2012) to our adopted distance. This gives SFR = 1.50$^{+0.39}_{-0.31}$ \msun\ yr$^{-1}$, consistent with the range implied by the calibrations of Rosa-Gonz\'{a}lez \etal (2002), and consistent with our HMXB-determined SFR at $\sim 1.5 \sigma$. Although not a highly significant difference, one explanation for this offset may be the incompleteness of this sample of HMXBs. G03 was derived for all binaries above $2 \times 10^{38}$ \es, whereas our completeness limit is $6 \times 10^{38}$ \es. A deeper observation of NGC 337 will complete the luminosity function down to lower limit used by G03, and allow for a more precise application of this correlation. Another possibility proposed by Shields \etal (2012) is that a significant discrepancy between SFR indicators and G03 may be seen if a galaxy is near the conclusion of a star burst epoch. In this scenario, HMXBs and ULXs have outlived classic SFR indicators, which have been attenuated as the stellar population ages. However, with a weak significance of only $1.5 \sigma$, the high abundance of HMXBs in NGC 337 is likely a statistical fluctuation rather than a galaxy observed in a special epoch. Nevertheless, NGC 337 hosts a rich population of ULX candidates and may prove valuable to the ongoing study of ultraluminous X-ray sources.

\begin{figure*}
\centering
{\setlength{\fboxrule}{1.5pt}
\fbox{\includegraphics[width=2.1in]{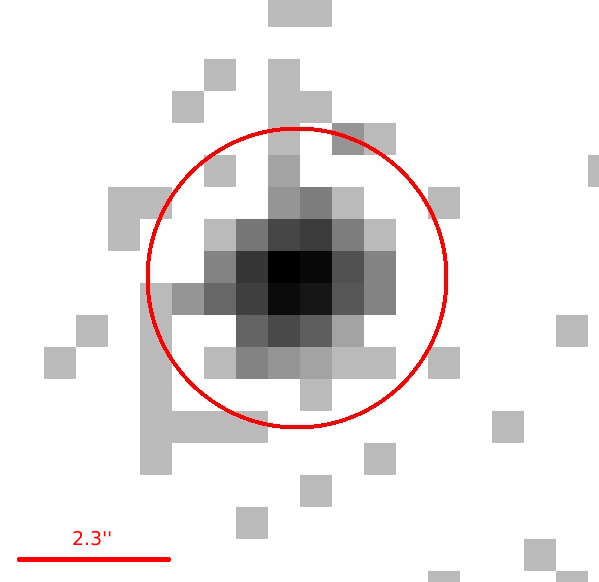}}
}
{\setlength{\fboxrule}{1.5pt}
\fbox{\includegraphics[width=2in]{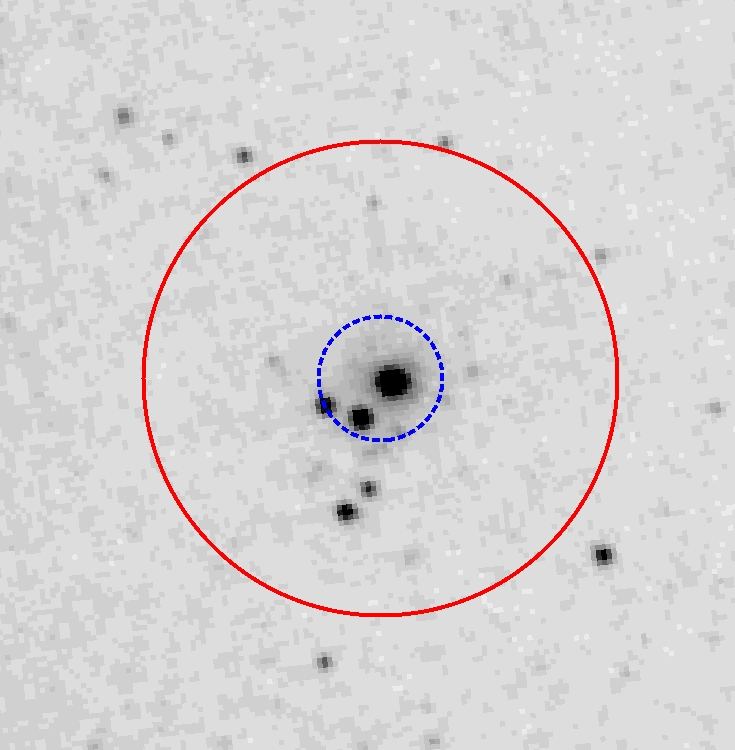}}
}
{\setlength{\fboxrule}{1.5pt}
\fbox{\includegraphics[width=2.05in]{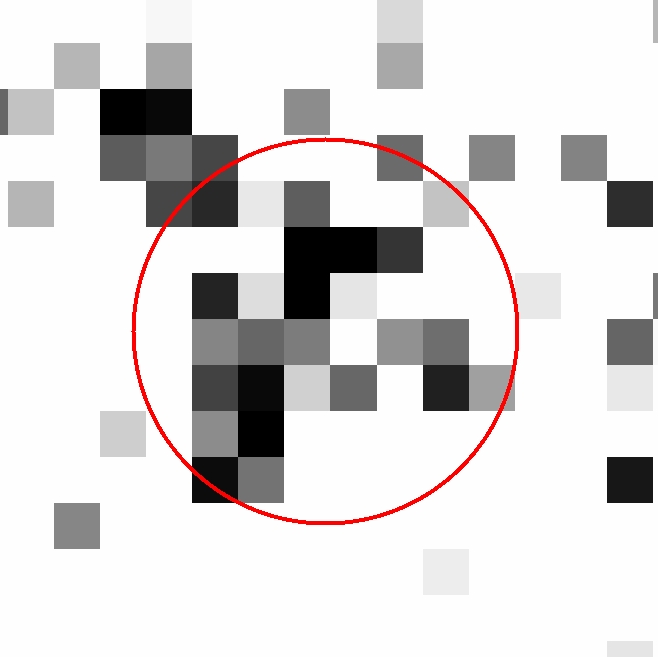}}
}

\caption{\textit{Left:} Unsmoothed image of ESO 501-23 X-1 from \emph{Chandra}. The red circle has a radius of 2.3$''$, equal to half the extraction circle. \textit{Middle:} HST WFC1 F606W image of X-1. The red circle has a radius of 2.3$''$, and the blue circle has a radius of 0.6$''$, equal to the \chandra 90\% source certainty region. \textit{Right:} H$\alpha$ image of X-1 from MDM 2.4m imaging.
}

\vspace{2mm}
\end{figure*}

     \subsection{ESO 501-23 X-1}

ESO 501-23 X-1 is located at R.A. = $\rm{10^h35^m22.2^s}$, Dec. = $\rm{-24^d45^m14^s}$. It is the brightest source discovered in this study, with a net count rate of $\approx$ 0.087 $\pm$ 0.003 cts s$^{-1}$ within a 2.3$''$ aperture. By inspection, we find a significant excess of photons above the background surrounding this extraction circle. We therefore include counts within 4.6$''$ of the source centroid, double the original extraction circle, and find a marginally higher rate of 0.090 $\pm$ 0.003 cts s$^{-1}$. The resulting spectrum is shown in Figure 5. We fit this spectrum with a power law including Galactic absorption (Kalberla \etal 2005), and a second variable absorption term. This produced a good fit to the spectrum ($\Gamma = 1.18^{+0.19}_{-0.11}$, N$\rm{_H}$ = 1.13$^{+7.07}_{-1.13} \times 10^{20}$ cm$^{-2}$, norm = 8.89$^{+1.90}_{-0.94} \times 10^{-5}$, \chired\ = 1.089 for 38 d.o.f.), and gives an unabsorbed flux of $1.08^{+0.05}_{-0.07} \times 10^{-12}$ \esc. Errors quoted are 90\% confidence intervals. Other simple models do not perform as well, but produce fluxes within $\sim$ 7\% of this value. CXB sources with a flux this large are rare, with M03 estimating 0.02 seen within the collective D25s of our sample. Although distance estimates vary considerably, they all imply an X-ray luminosity $> 5 \times 10^{39}$ \es\ for X-1. Considering its spectral shape, brightness, off-axis position, and coincidence with H$\alpha$ emission (Figure 4), this is likely a \emph{bona\ fide} ULX.

Power laws have been widely used to model ultraluminous X-ray sources in the literature and typically describe their spectra well, but lack the complexity to accurately reflect ULX spectra with sampling in excess of $\sim$ 10,000 counts (Gladstone \etal 2009). This demonstrates the need for high quality data if one plans to make accurate inferences about the source's physical nature. With only $\sim$ 800 counts, we cannot discriminate between more complicated spectral models, and thus do not attempt an analysis of this object in the context of the ULX progenitor debate. A deeper exposure with a higher effective area spectrograph, such as the \emph{EPIC} instrument aboard \emph{XMM-Newton}, could shed light on this promising object.

Although we do not speculate on the mass of X-1's putative black hole, we can use its spectral index to constrain its luminosity. Berghea \etal (2008) assembled a sample of ULXs to study their collective properties, showing in particular a negative correlation between spectral hardness and X-ray luminosity. The evidence for this relation was strengthened by Sutton \etal (2012), who reprocessed the Berghea  sample luminosities using best-fit power law spectral parameters. They found $<$$\Gamma$$>$ = 2.10 $\pm$ 0.07 for ULXs with L$_X < 10^{40}$ \es, and $<$$\Gamma$$>$ = 1.54 $\pm$ 0.06 for ULXs with L$_X > 10^{40}$ \es. If ESO 501-23 X-1 follows this pattern, its spectral index of $\Gamma \approx 1.2^{+0.2}_{-0.1}$ implies L$_X > 10^{40}$ \es. We can use this prior to evaluate distance estimates to ESO 501-23, which show scatter at the factor of 2 level. Tully (1988) and (2008) give 14.2 Mpc and 7.01 Mpc respectively, and Virgo infall-corrected redshift measurements give $\sim$ 13 Mpc (Mould \etal 2000), assuming standard cosmology (H$\rm{_0}$ = 70 km s$^{-1}$ Mpc$^{-1}$). The closest of these estimates gives an unabsorbed luminosity of L$_X = 6.4^{+0.3}_{-0.4} \times 10^{39}$ \es\ for X-1, securely in the ULX regime. However, if this is the correct distance to the galaxy, X-1 would be one of the hardest L$_X < 10^{40}$ \es\ ULXs yet discovered. We do not rule out this possibility, as there is a large intrinsic scatter in $\Gamma$ for dimmer ULXs (see Figure 6 in Sutton \etal 2012). However, the more distant estimates imply a luminosity of $\gtrsim 2 \times 10^{40}$ \es\ for X-1, a value much more consistent with the findings of Sutton \etal (2012). Although we cannot resolve this disagreement, our observations favor the larger distance calculations.

We searched archival HST data for a potential counterpart to this source. Only a F606W ACS/WFC image of the source position exists in the archive, shown in Figure 4. This image reveals multiple bright optical sources within the \chandra positional error circle, with a diffuse light occupying a radius of $\sim$ 100 parsecs. This may imply that X-1 is a collection of bright X-ray point sources that together reach an ultraluminous level. Further X-ray observations of X-1 could disfavor this interpretation if substantial variability is detected.  Alternatively, the brightest source in the HST image could be a high mass stellar companion of an accreting black hole or a tight stellar population in which the ULX resides, both of which have been claimed for other ULXs (e.g. Roberts \etal 2008, Farrell \etal 2012). Additionally, the diffuse light surrounding this object may be explained by a circum-ULX nebula, which has been reported in some cases (e.g. Pakull \& Mirioni 2003), or could simply be blended light from low mass stars. Future spectroscopic and color information could help discriminate between these possibilities, and add to statistics about ULX companions and environments.

\begin{figure}
\includegraphics[width=3.5in]{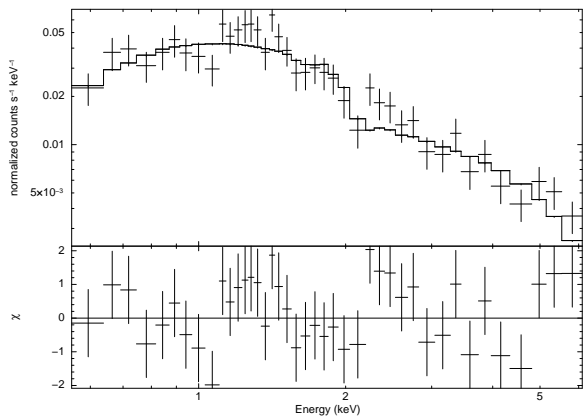}
\caption{ESO 501-23 X-1 spectrum extracted from a 4.6$''$ radius circle and binned to include at least 20 counts per tick. Also plotted is the best fit absorbed power law model described in $\S$4.2, and the $\chi^2$ residuals of each bin.}
\end{figure}

\section{Conclusions}
Drawing from the sample of W11, we select eight bulge-less disk galaxies for study using the \chandra X-ray Observatory. We report the discovery of several strong ULX candidates in the star forming galaxy NGC 337, which imply an SFR of 6.8$^{+4.4}_{-3.5}$ \msun\ yr$^{-1}$. This is consistent at $\sim$ 1.5$\sigma$ with the current best estimate of the star formation rate of NGC 337, and may be made more consistent with a higher signal-to-noise determination of its HMXB luminosity function.  We discovered a bright X-ray source in the galaxy ESO 501-23, and argue that it is a \emph{bona\ fide} ULX based on its flux, spectrum, off-center location, and coincidence with H$\alpha$. Its spectral index implies a luminosity $\gtrsim 10^{40}$ \es\ which, if correct, argues the distance to ESO 501-23 is larger than some previous published estimates. Further investigation of this source may reveal the nature of its optical counterpart, which is seen in HST imaging. Additional ULX candidates are identified in IC 1291, PGC 3853, NGC 5964, and NGC 2805, though deeper X-ray observations are required to adequately model their spectra. None of these galaxies appear active, confirming their optical classification in the literature, and supporting the notion that AGN in bulge-less spirals are rare. We also note that an X-ray source lies near the center of ESO 501-23, but do no claim that it is an AGN due to large uncertainties in the exact central coordinates of the galaxy. This source may warrant further attention in future X-ray studies of this galaxy. Finally, we do not detect the type II-P supernova SN 2011DQ, which exploded in NGC 337 prior to our observation.

\acknowledgments

Support for this work was provided by the National Aeronautics and Space Administration through Chandra Award Number GO1-12113X issued by the Chandra X-ray Observatory Center, which is operated by the Smithsonian Astrophysical Observatory for and on behalf of the National Aeronautics Space Administration under contract NAS8-03060. Funding for SDSS-III has been provided by the Alfred P. Sloan Foundation, the Participating Institutions, the National Science Foundation, and the U.S. Department of Energy Office of Science. The SDSS-III web site is http://www.sdss3.org/. SDSS-III is managed by the Astrophysical Research Consortium for the Participating Institutions of the SDSS-III Collaboration including the University of Arizona, the Brazilian Participation Group, Brookhaven National Laboratory, University of Cambridge, University of Florida, the French Participation Group, the German Participation Group, the Instituto de Astrofisica de Canarias, the Michigan State/Notre Dame/JINA Participation Group, Johns Hopkins University, Lawrence Berkeley National Laboratory, Max Planck Institute for Astrophysics, New Mexico State University, New York University, Ohio State University, Pennsylvania State University, University of Portsmouth, Princeton University, the Spanish Participation Group, University of Tokyo, University of Utah, Vanderbilt University, University of Virginia, University of Washington, and Yale University. 
This publication makes use of data products from the Wide-field Infrared Survey Explorer, which is a joint project of the University of California, Los Angeles, and the Jet Propulsion Laboratory/California Institute of Technology, funded by the National Aeronautics and Space Administration. This research has made use of the USNO Image and Catalogue Archive operated by the United States Naval Observatory, Flagstaff Station (http://www.nofs.navy.mil/data/fchpix/). GS would like to thank Dale Mudd for his helpful comments.

\clearpage

\begin{deluxetable*}{clrcccccccc}

\tablewidth{0pt}
\tablecaption{Galaxy Properties \label{tab:obs}}
\tablehead{
\colhead{Galaxy}
&\colhead{R.A.}
&\colhead{Dec.}
&\colhead{Error}
&\colhead{Dist.}
&\colhead{Ma.}
&\colhead{Mi.}
&\colhead{P.A.}
&\colhead{Type}
&\colhead{N$\rm{_H}$}
&\colhead{AGN} \\
\colhead{Name}
&\colhead{(J2000)}
&\colhead{(J2000)}
&\colhead{($''$)}
&\colhead{(Mpc)}
&\colhead{($''$)}
&\colhead{($''$)}
&\colhead{($^{\circ}$)}
&\colhead{(T)}
&\colhead{($\times 10^{20}$)}
&\colhead{Limit} \\
\colhead{(1)}
&\colhead{(2)}
&\colhead{(3)}
&\colhead{(4)}
&\colhead{(5)}
&\colhead{(6)}
&\colhead{(7)}
&\colhead{(8)}
&\colhead{(9)}
&\colhead{(10)}
&\colhead{(11)}
}
\startdata

NGC 337&00:59:50.09&$-$07:34:40.7&1.25&20.7 [T88]&173&109&118 $\pm$ 5&7.0&7.34&38.27 \\
IC 1291&18:33:52.57&+49:16:43.0&1.25&31.5 [T88]&109&91&131 $\pm$ 2&8.0&5.55&38.62 \\
PGC 3853&01:05:04.88&$-$06:12:44.6&1.25&11.4 [T08]&250&213&105.3 $\pm$ 0.2&7.0&5.11&37.92 \\
NGC 5964&15:37:36.32&+05:58:24.3&1.25&24.7 [T88]&250&194&136.7 $\pm$ 1.2&7.0&4.29&38.40 \\
UGC 6446&11:26:40.46&+53:44:48.3&0.50&18.0 [T08]&213&137&189.4 $\pm$ 0.5&7.0&0.97&38.09 \\
NGC 2805&09:20:20.41&+64:06:10.0&1.25&28.0 [T88]&379&287&300 $\pm$ 3&7.0&4.01&38.52 \\
ESO 501-23&10:35:23.20&$-$24:45:14.9&1.25&7.01 [T08]&208&165&224 $\pm$ 2&8.0&4.58&37.31 \\
NGC 3794&11:40:54.24&+56:12:07.2&0.50&19.2 [T08]&134&87&123.1 $\pm$ 1.1&6.5&0.92&38.15 \\
\enddata

\tablecomments{
Column 1: Galaxy name.
Columns 2, 3 and 4: R.A. and Dec. of galactic nucleus with uncertainties, taken from 2MASS, SDSS (UGC6446, NGC3794) and 6dF (ESO501-23). These are typically photometric centers; see W11 for rotational centers.
Column 5: Distance to the galaxy from Tully \etal (1988, [T88]) or (2008, T[08]) as indicated.
Columns 6 and 7: Semi-major and semi-minor axes of the 25th magnitude arcsec$^{-2}$ B-band isophote from de Vaucouleurs \etal (1991).
Column 8: Pitch angle of the galaxy, from Watson \etal (2012).
Column 9: De Vaucouleurs T Type.
Column 10: Galactic hydrogen column density per cm$^2$ along line-of-sight to galaxy from Kalberla \etal (2005).
Column 11: Log 2\sig\ upper limit for a putative AGN luminosity at the location from Columns 2 and 3.
}

\end{deluxetable*}


\begin{deluxetable*}{cccccccc}
\setlength{\tabcolsep}{0.15in}
\tablewidth{0pt}
\tablecaption{Observations \label{tab:obs}}
\tablehead{
\colhead{Galaxy}
&\colhead{R.A.}
&\colhead{Dec.}
&\colhead{Obs ID}
&\colhead{Observation}
&\colhead{Exposure}
&\colhead{Chip}
&\colhead{Filtered Time} \\
\colhead{}
&\colhead{}
&\colhead{}
&\colhead{}
&\colhead{Date}
&\colhead{}
&\colhead{}
&\colhead{} \\
\colhead{}
&\colhead{(J2000)}
&\colhead{(J2000)}
&\colhead{}
&\colhead{(yyyy-mm-dd)}
&\colhead{(ks)}
&\colhead{}
&\colhead{(ks)} \\
\colhead{(1)}
&\colhead{(2)}
&\colhead{(3)}
&\colhead{(4)}
&\colhead{(5)}
&\colhead{(6)}
&\colhead{(7)}
&\colhead{(8)}
}
\startdata
\\
NGC 337&00:59:50.1&$-$07:34:41&12979&2011-07-19&9.93&S2&9.61 \\
&&&&&&S3&9.93 \\
&&&&&&S4&N/A \\ \\
IC 1291&18:33:52.6&+49:16:43&12980&2011-04-06&9.93&S2&8.98 \\
&&&&&&S3&9.93 \\
&&&&&&S4&8.66 \\ \\
PGC 3853&01:05:04.9&$-$06:12:45&12981&2011-06-26&9.93&S2&9.47 \\
&&&&&&S3&9.81 \\
&&&&&&S4&9.79 \\ \\
NGC 5964&15:37:36.3&+05:58:24&12982&2011-04-19&9.93&S2&9.23 \\
&&&&&&S3&9.93 \\
&&&&&&S4&9.23 \\ \\
UGC 6446&11:26:40.4&+53:44:51&12983&2012-01-17&9.93&S2&8.98 \\
&&&&&&S3&9.93 \\
&&&&&&S4&N/A \\ \\
NGC 2805&09:20:20.4&+64:06:10&12984&2011-11-06&9.93&S2&8.91 \\
&&&&&&S3&9.61 \\
&&&&&&S4&8.66 \\ \\
ESO 501-23&10:35:23.2&$-$24:45:09&12985&2010-11-09&9.93&S2&8.82 \\
&&&&&&S3&9.93 \\
&&&&&&S4&9.61 \\ \\
NGC 3794&11:40:53.4&+56:12:07&12986&2012-01-18&9.93&S2&9.85 \\
&&&&&&S3&9.93 \\
&&&&&&S4&8.84 \\ \\

\enddata

\tablecomments{
Column 1: Galaxy name.
Columns 2 and 3: J2000 R.A. and Dec. of the pointing location of the observation.
Column 4: \chandra observation ID. 
Column 5: Date of observation. 
Column 6: Unfiltered observation time in kiloseconds. 
Columns 7 and 8: The respective time filter for each ACIS-S chip. N/A implies that no sources were detected on the chip, and thus no time filter was applied.
}
\end{deluxetable*}


\clearpage

\begin{landscape}
\begin{deluxetable}{cccccccccccc}
\setlength{\tabcolsep}{0.08in}
\tablewidth{0pt}
\tablecaption{Full Source List}
\tablehead{
\colhead{Obs ID}
&\colhead{Chip}
&\colhead{ID}
&\colhead{Source}
&\colhead{R.A.}
&\colhead{Decl.}
&\colhead{A$\rm{_{eff}}$}
&\colhead{$\Delta$}
&\colhead{Net Counts}
&\colhead{Soft Counts}
&\colhead{Hard Counts}
&\colhead{Inside} \\
\colhead{}
&\colhead{Num}
&\colhead{Num}
&\colhead{Name}
&\colhead{}
&\colhead{}
&\colhead{}
&\colhead{}
&\colhead{$\pm$ Error}
&\colhead{$\pm$ Error}
&\colhead{$\pm$ Error}
&\colhead{D25?} \\
\colhead{}
&\colhead{}
&\colhead{}
&\colhead{(CXOU)}
&\colhead{(J2000)}
&\colhead{(J2000)}
&\colhead{(cm$^2$)}
&\colhead{($'$)}
&\colhead{(0.3-8 keV)}
&\colhead{(0.3-2 keV)}
&\colhead{(2-8 keV)}
&\colhead{} \\
\colhead{(1)}
&\colhead{(2)}
&\colhead{(3)}
&\colhead{(4)}
&\colhead{(5)} 
&\colhead{(6)}
&\colhead{(7)}
&\colhead{(8)}
&\colhead{(9)}
&\colhead{(10)}
&\colhead{(11)}
&\colhead{(12)}
}
\startdata 

\textbf{12979}&\textbf{S2}&1&J005951.7$-$073052&00 59 51.7&$-$07 30 52&589&3.7&14.7 $\pm$ 3.9&14.0 $\pm$ 3.7&0.7 $\pm$ 1.0&...   \\
&&2&J005951.5$-$073134&00 59 51.5&$-$07 31 34&597&3.0&10.8 $\pm$ 3.3&4.0 $\pm$ 2.0&6.9 $\pm$ 2.6&...   \\
&&3&J005944.6$-$073058&00 59 44.6&$-$07 30 58&587&3.7&8.8 $\pm$ 3.0&3.9 $\pm$ 2.0&4.9 $\pm$ 2.2&...   \\
&&4&J005937.3$-$073034&00 59 37.3&$-$07 30 34&486&4.9&9.2 $\pm$ 3.2&3.7 $\pm$ 2.0&5.6 $\pm$ 2.5&...   \\ 
&\textbf{S3}&5&J005952.3$-$073447&00 59 52.3&$-$07 34 47&628&0.8&43.8 $\pm$ 6.7&32.3 $\pm$ 5.7&11.5 $\pm$ 3.5&Y   \\
&&6&J005951.9$-$073458&00 59 51.9&$-$07 34 58&589&0.8&14.0 $\pm$ 3.9&4.5 $\pm$ 2.2&9.5 $\pm$ 3.2&Y   \\
&&7&J005950.6$-$073458&00 59 50.6&$-$07 34 58&627&0.6&315.4 $\pm$ 17.8&194.8 $\pm$ 14.0&120.6 $\pm$ 11.0&Y   \\
&&8&J005950.4$-$073454&00 59 50.4&$-$07 34 54&627&0.5&43.2 $\pm$ 6.7&26.7 $\pm$ 5.3&16.5 $\pm$ 4.1&Y   \\
\enddata
\end{deluxetable}
\clearpage


\begin{deluxetable}{cccccccccccc}
\addtocounter{table}{-1}
\setlength{\tabcolsep}{0.08in}
\tablewidth{0pt}
\tablecaption{Continued}
\tablehead{
\colhead{Obs ID}
&\colhead{Chip}
&\colhead{ID}
&\colhead{Source}
&\colhead{R.A.}
&\colhead{Decl.}
&\colhead{A$\rm{_{eff}}$}
&\colhead{$\Delta$}
&\colhead{Net Counts}
&\colhead{Soft Counts}
&\colhead{Hard Counts}
&\colhead{Inside} \\
\colhead{}
&\colhead{Num}
&\colhead{Num}
&\colhead{Name}
&\colhead{}
&\colhead{}
&\colhead{}
&\colhead{}
&\colhead{$\pm$ Error}
&\colhead{$\pm$ Error}
&\colhead{$\pm$ Error}
&\colhead{D25?} \\
\colhead{}
&\colhead{}
&\colhead{}
&\colhead{(CXOU)}
&\colhead{(J2000)}
&\colhead{(J2000)}
&\colhead{(cm$^2$)}
&\colhead{($'$)}
&\colhead{(0.3-8 keV)}
&\colhead{(0.3-2 keV)}
&\colhead{(2-8 keV)}
&\colhead{} \\
\colhead{(1)}
&\colhead{(2)}
&\colhead{(3)}
&\colhead{(4)}
&\colhead{(5)} 
&\colhead{(6)}
&\colhead{(7)}
&\colhead{(8)}
&\colhead{(9)}
&\colhead{(10)}
&\colhead{(11)}
&\colhead{(12)}
}
\startdata 
\textbf{12981}&\textbf{S2}&48&J010509.0$-$060650&01 05 09.0&$-$06 06 50&559&5.9&21.4 $\pm$ 4.7&16.9 $\pm$ 4.1&4.5 $\pm$ 2.2&...   \\
&&49&J010454.3$-$060622&01 04 54.3&$-$06 06 22&553&6.7&6.1 $\pm$ 2.8&3.4 $\pm$ 2.0&2.7 $\pm$ 2.0&...   \\
&&50&J010443.6$-$060808&01 04 43.6&$-$06 08 08&546&6.7&31.4 $\pm$ 5.8&22.5 $\pm$ 4.8&8.9 $\pm$ 3.2&...   \\
&&51&J010442.2$-$060712&01 04 42.2&$-$06 07 12&536&7.6&28.7 $\pm$ 5.5&19.5 $\pm$ 4.5&9.2 $\pm$ 3.2&...   \\ 
&\textbf{S3}&52&J010511.3$-$061207&01 05 11.3&$-$06 12 07&652&1.9&182.6 $\pm$ 13.5&130.8 $\pm$ 11.4&51.8 $\pm$ 7.2&Y   \\
&&53&J010506.5$-$061509&01 05 06.5&$-$06 15 09&567&2.6&21.0 $\pm$ 4.6&16.0 $\pm$ 4.0&5.0 $\pm$ 2.2&...   \\
&&54&J010502.8$-$061536&01 05 02.8&$-$06 15 36&598&3.0&18.5 $\pm$ 4.4&14.7 $\pm$ 3.9&3.9 $\pm$ 2.0&...   \\
&&55&J010502.6$-$061509&01 05 02.6&$-$06 15 09&642&2.6&12.7 $\pm$ 3.6&8.9 $\pm$ 3.0&3.9 $\pm$ 2.0&...   \\
\enddata
\end{deluxetable}
\clearpage


\begin{deluxetable}{cccccccccccc}
\addtocounter{table}{-1}
\setlength{\tabcolsep}{0.08in}
\tablewidth{0pt}
\tablecaption{Continued}
\tablehead{
\colhead{Obs ID}
&\colhead{Chip}
&\colhead{ID}
&\colhead{Source}
&\colhead{R.A.}
&\colhead{Decl.}
&\colhead{A$\rm{_{eff}}$}
&\colhead{$\Delta$}
&\colhead{Net Counts}
&\colhead{Soft Counts}
&\colhead{Hard Counts}
&\colhead{Inside} \\
\colhead{}
&\colhead{Num}
&\colhead{Num}
&\colhead{Name}
&\colhead{}
&\colhead{}
&\colhead{}
&\colhead{}
&\colhead{$\pm$ Error}
&\colhead{$\pm$ Error}
&\colhead{$\pm$ Error}
&\colhead{D25?} \\
\colhead{}
&\colhead{}
&\colhead{}
&\colhead{(CXOU)}
&\colhead{(J2000)}
&\colhead{(J2000)}
&\colhead{(cm$^2$)}
&\colhead{($'$)}
&\colhead{(0.3-8 keV)}
&\colhead{(0.3-2 keV)}
&\colhead{(2-8 keV)}
&\colhead{} \\
\colhead{(1)}
&\colhead{(2)}
&\colhead{(3)}
&\colhead{(4)}
&\colhead{(5)} 
&\colhead{(6)}
&\colhead{(7)}
&\colhead{(8)}
&\colhead{(9)}
&\colhead{(10)}
&\colhead{(11)}
&\colhead{(12)}
}
\startdata 
\textbf{12983}&\textbf{S2}&94&J112646.2$+$534921&11 26 46.2&$+$53 49 21&555&4.6&65.2 $\pm$ 8.1&40.8 $\pm$ 6.4&24.4 $\pm$ 5.0&...   \\
&&95&J112622.4$+$534722&11 26 22.4&$+$53 47 22&586&3.5&9.8 $\pm$ 3.2&6.0 $\pm$ 2.4&3.8 $\pm$ 2.0&...   \\
&&96&J112632.0$+$535022&11 26 32.0&$+$53 50 22&522&5.6&9.4 $\pm$ 3.2&7.9 $\pm$ 2.8&1.5 $\pm$ 1.4&...   \\
&&97&J112629.1$+$535007&11 26 29.1&$+$53 50 07&469&5.4&8.5 $\pm$ 3.0&6.9 $\pm$ 2.6&1.7 $\pm$ 1.4&...   \\
&&98&J112626.1$+$534955&11 26 26.1&$+$53 49 55&479&5.4&9.6 $\pm$ 3.2&9.9 $\pm$ 3.2&-0.3 $\pm$ 0.1&...   \\
&&99&J112623.5$+$534855&11 26 23.5&$+$53 48 55&542&4.7&7.6 $\pm$ 2.8&5.0 $\pm$ 2.2&2.7 $\pm$ 1.7&...   \\
&&100&J112606.4$+$534857&11 26 06.4&$+$53 48 57&434&6.3&18.7 $\pm$ 4.5&10.5 $\pm$ 3.3&8.2 $\pm$ 3.0&...   \\ 
&\textbf{S3}&101&J112656.1$+$533932&11 26 56.1&$+$53 39 32&612&5.9&24.6 $\pm$ 5.1&18.5 $\pm$ 4.4&6.1 $\pm$ 2.7&...   \\
\enddata
\end{deluxetable}
\clearpage


\begin{deluxetable}{cccccccccccc}
\addtocounter{table}{-1}
\setlength{\tabcolsep}{0.08in}
\tablewidth{0pt}
\tablecaption{Continued}
\tablehead{
\colhead{Obs ID}
&\colhead{Chip}
&\colhead{ID}
&\colhead{Source}
&\colhead{R.A.}
&\colhead{Decl.}
&\colhead{A$\rm{_{eff}}$}
&\colhead{$\Delta$}
&\colhead{Net Counts}
&\colhead{Soft Counts}
&\colhead{Hard Counts}
&\colhead{Inside} \\
\colhead{}
&\colhead{Num}
&\colhead{Num}
&\colhead{Name}
&\colhead{}
&\colhead{}
&\colhead{}
&\colhead{}
&\colhead{$\pm$ Error}
&\colhead{$\pm$ Error}
&\colhead{$\pm$ Error}
&\colhead{D25?} \\
\colhead{}
&\colhead{}
&\colhead{}
&\colhead{(CXOU)}
&\colhead{(J2000)}
&\colhead{(J2000)}
&\colhead{(cm$^2$)}
&\colhead{($'$)}
&\colhead{(0.3-8 keV)}
&\colhead{(0.3-2 keV)}
&\colhead{(2-8 keV)}
&\colhead{} \\
\colhead{(1)}
&\colhead{(2)}
&\colhead{(3)}
&\colhead{(4)}
&\colhead{(5)} 
&\colhead{(6)}
&\colhead{(7)}
&\colhead{(8)}
&\colhead{(9)}
&\colhead{(10)}
&\colhead{(11)}
&\colhead{(12)}
}
\startdata 
\textbf{12985}&\textbf{S2}&136&J103535.3$-$244318&10 35 35.3&$-$24 43 18&563&3.3&26.9 $\pm$ 5.2&19.9 $\pm$ 4.5&7.0 $\pm$ 2.6&...   \\
&&137&J103510.4$-$243846&10 35 10.4&$-$24 38 46&482&6.7&47.8 $\pm$ 7.0&27.7 $\pm$ 5.3&20.1 $\pm$ 4.6&...   \\
&&138&J103514.5$-$243939&10 35 14.5&$-$24 39 39&562&5.6&6.8 $\pm$ 2.6&4.9 $\pm$ 2.2&1.9 $\pm$ 1.4&...   \\
&&139&J103507.0$-$244149&10 35 07.0&$-$24 41 49&549&4.7&7.6 $\pm$ 2.8&3.9 $\pm$ 2.0&3.7 $\pm$ 2.0&...   \\
&&140&J103520.7$-$243708&10 35 20.7&$-$24 37 08&540&7.8&14.4 $\pm$ 4.0&11.5 $\pm$ 3.5&2.9 $\pm$ 2.0&...   \\ 

&\textbf{S3}&141&J103522.2$-$244513&10 35 22.2&$-$24 45 13&653&0.3&867.0 $\pm$ 29.5&528.7 $\pm$ 23.0&338.3 $\pm$ 18.4&Y   \\
&&142&J103523.3$-$244517&10 35 23.3&$-$24 45 17&654&0.4&68.3 $\pm$ 8.3&40.7 $\pm$ 6.4&27.6 $\pm$ 5.3&Y   \\
&&143&J103526.9$-$244745&10 35 26.9&$-$24 47 45&602&3.0&7.6 $\pm$ 2.8&6.7 $\pm$ 2.6&0.9 $\pm$ 1.0&...   \\
\enddata

\tablecomments{
Column 1: The \chandra observation ID of the exposure that found this source, as well as the target galaxy of that observation. 
Column 2: The ACIS-S chip that this source was observed on.
Column 3: Assigned ID number of the source. 
Column 4: Chandra ID name.
Columns 5, 6: R.A. and Decl. of source in the J2000 epoch.
Column 7: The 1.7 keV effective area at the location of the source. This is the net instrument response, integrated over the dither path of the observation. Sources with lower effective areas are typically on CCD chip edges.
Column 8: Angular distance from the pointing location of the observation to the source, in arcminutes.
Columns 9, 10, 11: Background subtracted net counts, with 1\sig\ uncertainties, in broad (0.3-8 keV), soft (0.3-2 keV) and hard (2-8 keV) bands respectively.
Column 12: Is this object within the D25 isophote of the target galaxy? ``Y'' if yes, ``...'' if no.
}

\end{deluxetable}

\clearpage

\begin{deluxetable*}{ccccccccccccc}
\setlength{\tabcolsep}{0.065in}
\tablewidth{0pt}
\tablecaption{USNO$-$B1.0 Cross Correlation}
\tablehead{
\colhead{N}
&\colhead{Source}
&\colhead{Identification}
&\colhead{$\Delta$}
&\colhead{B1}
&\colhead{R1}
&\colhead{B2}
&\colhead{R2}
&\colhead{I}
&\colhead{R.A.}
&\colhead{Decl.}
&\colhead{$\alpha$}
&\colhead{$\delta$} \\
\colhead{}
&\colhead{Name}
&\colhead{(USNO$-$B1.0)}
&\colhead{($''$)}
&\colhead{(mag)}
&\colhead{(mag)}
&\colhead{(mag)}
&\colhead{(mag)}
&\colhead{(mag)}
&\colhead{(B1.0)}
&\colhead{(B1.0)}
&\colhead{(mas/yr)}
&\colhead{(mas/yr)} \\
\colhead{(1)}
&\colhead{(2)}
&\colhead{(3)}
&\colhead{(4)}
&\colhead{(5)}
&\colhead{(6)}
&\colhead{(7)}
&\colhead{(8)}
&\colhead{(9)}
&\colhead{(10)}
&\colhead{(11)}
&\colhead{(12)}
&\colhead{(13)}
}
\startdata

1&79.6.01&0824$-$0013640&0.68&19.1&19.31&19.76&19.65&18.61&00 59 51.705&$-$07 30 52.15&0&0 \\
12&79.7.09&0824$-$0013616&1.06& &18.26& & &16.17&00 59 47.553&$-$07 34 15.91&0&0 \\
13&79.7.10&0824$-$0013575&0.12&21.3&19.3&21.81&20.14&18.7&00 59 40.451&$-$07 33 47.90&0&0 \\
14&79.7.11&0824$-$0013656&1.46&17.81& & &18.66&18.59&00 59 53.323&$-$07 35 22.20&$-$62&4 \\
16&79.7.13&0823$-$0014391&1.66&15.88&15.48&15.89&14.37&13.61&00 59 48.055&$-$07 38 06.60&4&$-$4 \\
20&79.7.17&0823$-$0014409&0.9& & & &20.32&18.71&00 59 56.411&$-$07 38 32.72&0&0 \\
21&79.7.19&0823$-$0014364&0.83&20.6& &20.83&20.31& &00 59 35.397&$-$07 36 21.56&0&0 \\
31&80.7.01&1393$-$0289816&0.13&20.2&19.56&20.98&20.17& &18 34 02.167&$+$49 18 17.09&0&0 \\
\enddata

\tablecomments{
Column 1: Source number from Table 4.
Column 2: Observation ID, CCD chip number, and \wave\ source number.
Column 3: Identification number for the identified USNO$-$B1.0 source.
Column 4: Offset between X-ray source and USNO$-$B1.0 source centroids.
Column 5-9: First and second B and R magnitudes, and I magnitudes, for each source. Typical errors are 0.3 mag.
Column 10-11: R.A. and Decl. of the source listed in USNO$-$B1.0 catalog.
Column 12-13: Proper motion of the source.
}

\end{deluxetable*}

\clearpage

\begin{deluxetable*}{cccllllllccc}
\setlength{\tabcolsep}{0.065in}
\tablewidth{0pt}
\tablecaption{SDSS DR8 Cross Correlation}
\tablehead{
\colhead{N}
&\colhead{Source}
&\colhead{Identification}
&\colhead{$\Delta$}
&\colhead{u}
&\colhead{g}
&\colhead{r}
&\colhead{i}
&\colhead{z}
&\colhead{R.A.}
&\colhead{Decl.}
&\colhead{cl} \\
\colhead{}
&\colhead{Name}
&\colhead{(SDSS DR8)}
&\colhead{($''$)}
&\colhead{(mag)}
&\colhead{(mag)}
&\colhead{(mag)}
&\colhead{(mag)}
&\colhead{(mag)}
&\colhead{(SDSS)}
&\colhead{(SDSS)}
&\colhead{} \\
\colhead{(1)}
&\colhead{(2)}
&\colhead{(3)}
&\colhead{(4)}
&\colhead{(5)}
&\colhead{(6)}
&\colhead{(7)}
&\colhead{(8)}
&\colhead{(9)}
&\colhead{(10)}
&\colhead{(11)}
&\colhead{(12)}
}
\startdata
  1&79.6.01&J005951.73$-$073052.1&0.325&20.064 ($\pm$0.054)&19.807 ($\pm$0.016)&19.523 ($\pm$0.016)&19.503 ($\pm$0.02) &19.316 ($\pm$0.06) &00 59 51.734&$-$07 30 52.16&6 \\
  9&79.7.06&J005949.48$-$073522.8&0.584&16.471 ($\pm$0.068)&14.233 ($\pm$0.004)&13.956 ($\pm$0.005)&13.744 ($\pm$0.006)&14.041 ($\pm$0.026)&00 59 49.482&$-$07 35 22.89&3 \\
 12&79.7.09&J005947.49$-$073416.0&0.556&20.181 ($\pm$0.075)&20.101 ($\pm$0.028)&20.305 ($\pm$0.039)&20.833 ($\pm$0.078)&21.287 ($\pm$0.41) &00 59 47.499&$-$07 34 16.02&3 \\
 13&79.7.10&J005940.46$-$073348.1&0.308&23.009 ($\pm$0.735)&21.504 ($\pm$0.076)&20.189 ($\pm$0.037)&19.648 ($\pm$0.033)&19.225 ($\pm$0.087)&00 59 40.464&$-$07 33 48.19&3 \\
 16&79.7.13&J005948.10$-$073808.1&0.107&19.459 ($\pm$0.037)&16.958 ($\pm$0.004)&15.761 ($\pm$0.004)&15.215 ($\pm$0.004)&14.875 ($\pm$0.005)&00 59 48.103&$-$07 38 08.19&6 \\
 17&79.7.14&J005954.59$-$073656.2&0.548&22.646 ($\pm$0.649)&23.19  ($\pm$0.314)&22.006 ($\pm$0.185)&21.437 ($\pm$0.168)&21.635 ($\pm$0.776)&00 59 54.590&$-$07 36 56.25&3 \\
 20&79.7.17&J005956.43$-$073832.3&0.495&22.734 ($\pm$0.685)&21.67  ($\pm$0.083)&20.404 ($\pm$0.046)&19.815 ($\pm$0.041)&19.37  ($\pm$0.112)&00 59 56.433&$-$07 38 32.36&3 \\
 21&79.7.19&J005935.40$-$073621.5&0.872&25.23  ($\pm$1.213)&20.934 ($\pm$0.033)&20.492 ($\pm$0.034)&20.394 ($\pm$0.043)&20.055 ($\pm$0.132)&00 59 35.404&$-$07 36 21.56&6 \\
\enddata

\end{deluxetable*}

\begin{deluxetable*}{cccllllllccc}
\setlength{\tabcolsep}{0.065in}
\addtocounter{table}{-1}
\tablewidth{0pt}
\tablecaption{Continued}
\tablehead{
\colhead{N}
&\colhead{Source}
&\colhead{Identification}
&\colhead{$\Delta$}
&\colhead{u}
&\colhead{g}
&\colhead{r}
&\colhead{i}
&\colhead{z}
&\colhead{R.A.}
&\colhead{Decl.}
&\colhead{cl} \\
\colhead{}
&\colhead{Name}
&\colhead{(SDSS DR8)}
&\colhead{($''$)}
&\colhead{(mag)}
&\colhead{(mag)}
&\colhead{(mag)}
&\colhead{(mag)}
&\colhead{(mag)}
&\colhead{(SDSS)}
&\colhead{(SDSS)}
&\colhead{} \\
\colhead{(1)}
&\colhead{(2)}
&\colhead{(3)}
&\colhead{(4)}
&\colhead{(5)}
&\colhead{(6)}
&\colhead{(7)}
&\colhead{(8)}
&\colhead{(9)}
&\colhead{(10)}
&\colhead{(11)}
&\colhead{(12)}
}
\startdata

133&84.8.03&J091955.38$+$635934.5&0.135&23.64  ($\pm$0.774)&22.558 ($\pm$0.111)&21.916 ($\pm$0.15) &21.392 ($\pm$0.105)&20.792 ($\pm$0.268)&09 19 55.386&$+$63 59 34.54&3 \\
134&84.8.04&J092004.93$+$635526.3&1.557&20.015 ($\pm$0.044)&19.765 ($\pm$0.014)&19.441 ($\pm$0.015)&19.416 ($\pm$0.02) &19.391 ($\pm$0.073)&09 20 04.939&$+$63 55 26.37&6 \\
154&86.6.02&J114109.68$+$561523.3&1.115&23.238 ($\pm$0.603)&22.78  ($\pm$0.189)&22.084 ($\pm$0.156)&21.101 ($\pm$0.108)&20.482 ($\pm$0.217)&11 41 09.680&$+$56 15 23.34&3 \\
160&86.7.02&J114047.66$+$561059.0&0.434&27.663 ($\pm$0.949)&29.494 ($\pm$0.203)&19.648 ($\pm$0.119)&20.029 ($\pm$0.287)&20.685 ($\pm$1.984)&11 40 47.665&$+$56 10 59.06&3 \\
163&86.7.06&J114112.36$+$561441.2&0.326&22.09  ($\pm$0.212)&22.112 ($\pm$0.095)&21.51  ($\pm$0.084)&21.329 ($\pm$0.116)&21.137 ($\pm$0.335)&11 41 12.362&$+$56 14 41.20&6 \\
164&86.7.07&J114109.64$+$560818.9&0.362&22.11  ($\pm$0.215)&19.829 ($\pm$0.016)&18.446 ($\pm$0.008)&17.227 ($\pm$0.006)&16.498 ($\pm$0.009)&11 41 09.643&$+$56 08 18.97&6 \\
165&86.7.08&J114104.40$+$561237.6&0.132&22.119 ($\pm$0.219)&22.832 ($\pm$0.174)&22.833 ($\pm$0.26) &22.871 ($\pm$0.433)&21.866 ($\pm$0.58) &11 41 04.409&$+$56 12 37.68&6 \\
166&86.7.10&J114035.56$+$560914.6&0.346&21.404 ($\pm$0.121)&21.242 ($\pm$0.046)&21.513 ($\pm$0.08) &21.411 ($\pm$0.122)&21.102 ($\pm$0.332)&11 40 35.563&$+$56 09 14.67&6 \\
\enddata

\tablecomments{
This table is the result of a cross-correlation between Table 3 and the Sloan Digital Sky Survey Data Release 8 point source catalog. IC 1291 and ESO 501-23 are not in the Sloan footprint, so none of these sources have SDSS counterparts.
Column 1: Source number from Table 4.
Column 2: Obseravation ID, CCD chip number, and \wave\ source number.
Column 3: Identification number for the identified SDSS DR8 source.
Column 4: Offset between centroid of \chandra and SDSS DR8 sources.
Columns 5-9: Magnitudes for each of the five Sloan Digital Sky Survey band passes (ugriz), with errors.
Columns 10-11: R.A. and Decl. of the SDSS DR8 source, in the J2000 epoch.
Column 12: Object class listed in the SDSS DR8 catalog. 3 stands for galaxy, whereas 6 stands for star. 
}

\end{deluxetable*}

\begin{deluxetable*}{rlccrrcccc}
\setlength{\tabcolsep}{0.062in}
\tablewidth{0pt}
\centering
\tablecaption{WISE Cross Correlation}
\tablehead{
\colhead{N}
&\colhead{Source}
&\colhead{Identification}
&\colhead{$\Delta$}
&\colhead{W1}
&\colhead{W2}
&\colhead{W3}
&\colhead{W4}
&\colhead{R.A.}
&\colhead{Decl.} \\
\colhead{}
&\colhead{Name}
&\colhead{(WISE)}
&\colhead{($''$)}
&\colhead{(mag)}
&\colhead{(mag)}
&\colhead{(mag)}
&\colhead{(mag)}
&\colhead{(WISE)}
&\colhead{(WISE)} \\
\colhead{(1)}
&\colhead{(2)}
&\colhead{(3)}
&\colhead{(4)}
&\colhead{(5)}
&\colhead{(6)}
&\colhead{(7)}
&\colhead{(8)}
&\colhead{(9)}
&\colhead{(10)}
}
\startdata
  1&79.6.01 &J005951.71$-$073052.0&0.476&15.564 ($\pm$0.053)&14.223 ($\pm$0.068)&12.155 ($\pm$0.459)&   $>$8.73   &00 59 51.718&$-$07 30 52.08 \\
  3&79.6.03 &J005944.62$-$073059.1&1.766&17.297 ($\pm$0.212)&16.096 ($\pm$0.308)&12.096 ($\pm$0.461)&   $>$8.169  &00 59 44.623&$-$07 30 59.13 \\
  7&79.7.04 &J005950.64$-$073457.1&1.525&11.578 ($\pm$0.022)&11.277 ($\pm$0.023)&$\ $6.762 ($\pm$0.015)&3.742 ($\pm$0.019)&00 59 50.646&$-$07 34 57.17 \\
 13&79.7.10&J005940.44$-$073347.9&0.243&15.838 ($\pm$0.068)&15.163 ($\pm$0.136)&   $>$11.97   &   $>$8.677  &00 59 40.444&$-$07 33 47.97 \\
 16&79.7.13&J005948.05$-$073806.6&1.603&12.231 ($\pm$0.024)&12.246 ($\pm$0.028)&   $>$11.296  &   $>$8.047  &00 59 48.059&$-$07 38 06.64 \\
 17&79.7.14&J005954.59$-$073656.3&0.621&16.560 ($\pm$0.112)&15.736 ($\pm$0.208)&11.916 ($\pm$0.374)&   $>$8.813  &00 59 54.590&$-$07 36 56.34 \\
 19&79.7.16&J010007.40$-$073521.9&0.799&16.751 ($\pm$0.129)&15.577 ($\pm$0.187)&   $>$11.541  &   $>$8.065  &01 00 07.400&$-$07 35 21.97 \\
 20&79.7.17&J005956.49$-$073832.0&1.054&15.665 ($\pm$0.057)&14.760 ($\pm$0.102)&11.776 ($\pm$0.330)&   $>$8.81   &00 59 56.500&$-$07 38 32.03 \\
\enddata
\end{deluxetable*}

\begin{deluxetable*}{rlccrrcccc}
\setlength{\tabcolsep}{0.062in}
\addtocounter{table}{-1}
\tablewidth{0pt}
\centering
\tablecaption{Continued}
\tablehead{
\colhead{N}
&\colhead{Source}
&\colhead{Identification}
&\colhead{$\Delta$}
&\colhead{W1}
&\colhead{W2}
&\colhead{W3}
&\colhead{W4}
&\colhead{R.A.}
&\colhead{Decl.} \\
\colhead{}
&\colhead{Name}
&\colhead{(WISE)}
&\colhead{($''$)}
&\colhead{(mag)}
&\colhead{(mag)}
&\colhead{(mag)}
&\colhead{(mag)}
&\colhead{(WISE)}
&\colhead{(WISE)} \\
\colhead{(1)}
&\colhead{(2)}
&\colhead{(3)}
&\colhead{(4)}
&\colhead{(5)}
&\colhead{(6)}
&\colhead{(7)}
&\colhead{(8)}
&\colhead{(9)}
&\colhead{(10)}
}
\startdata

 58&81.7.08 &J010516.06$-$061747.9&0.680&16.279 ($\pm$0.069)&15.404 ($\pm$0.128)&   $>$11.867  &   $>$9.251  &01 05 16.062&$-$06 17 47.98 \\
 59&81.7.09 &J010500.17$-$061540.4&0.167&17.193 ($\pm$0.143)&16.061 ($\pm$0.206)&   $>$12.589  &   $>$9.284  &01 05 00.177&$-$06 15 40.48 \\
 62&81.7.12&J010515.54$-$061640.2&1.202&16.474 ($\pm$0.079)&15.746 ($\pm$0.165)&12.295 ($\pm$0.397)&   $>$8.862  &01 05 15.546&$-$06 16 40.29 \\
 64&81.7.14&J010517.06$-$061825.1&0.938&17.810 ($\pm$0.245)&15.711 ($\pm$0.158)&12.568 ($\pm$0.529)&   $>$8.548  &01 05 17.063&$-$06 18 25.15 \\
 67&81.8.01&J010529.68$-$061733.4&0.629&17.052 ($\pm$0.137)&16.000 ($\pm$0.225)&12.240 ($\pm$0.402)&   $>$8.813  &01 05 29.680&$-$06 17 33.49 \\
 69&81.8.03&J010513.11$-$062152.3&1.678&16.444 ($\pm$0.082)&15.648 ($\pm$0.165)&   $>$12.178  &   $>$9.081  &01 05 13.116&$-$06 21 52.30 \\
 73&82.6.04&J153731.62$+$060316.0&0.142&17.175 ($\pm$0.173)&15.833 ($\pm$0.183)&   $>$12.771  &   $>$8.659  &15 37 31.621&$+$06 03 16.09 \\
 74&82.6.05&J153724.64$+$060102.9&0.326&17.298 ($\pm$0.198)&16.810 ($\pm$0.459)&12.542 ($\pm$0.438)&   $>$8.794  &15 37 24.644&$+$06 01 02.94 \\
\enddata

\tablecomments{
Column 1: Source number from Table 4.
Column 2: Observation ID, CCD chip number, and \wave\ source number.
Column 3: Identification number for the identified WISE source.
Column 4: Offset between centroid of \chandra and WISE sources.
Columns 5-8: W1, W2, W3 and W4 WISE magnitudes respectively, with errors. A $>$ implies a magnitude lower limit.
Columns 9-10: R.A. and Decl. of the WISE source, in the J2000 epoch.
}

\end{deluxetable*}

\clearpage

\begin{deluxetable*}{lccccccccllcc}
\setlength{\tabcolsep}{0.065in}
\tiny
\tablewidth{0pt}
\tablecaption{Sources in the Sample Galaxies}
\tablehead{
\colhead{Source}
&\colhead{N}
&\colhead{\emph{Chandra} ID}
&\colhead{D}
&\colhead{Net}
&\colhead{Net}
&\colhead{Net}
&\colhead{Hardness}
&\colhead{$\Gamma$}
&\colhead{Unabs.}
&\colhead{Inferred}
&\colhead{H$\rm{\alpha}$?}
&\colhead{Notes} \\
\colhead{Name}
&\colhead{}
&\colhead{(CXOU)}
&\colhead{($''$)}
&\colhead{Counts}
&\colhead{Soft}
&\colhead{Hard}
&\colhead{Ratio}
&\colhead{}
&\colhead{Flux}
&\colhead{L$_X$}
&\colhead{}
&\colhead{} \\
\colhead{(1)}
&\colhead{(2)}
&\colhead{(3)}
&\colhead{(4)}
&\colhead{(5)} 
&\colhead{(6)}
&\colhead{(7)}
&\colhead{(8)}
&\colhead{(9)}
&\colhead{(10)}
&\colhead{(11)}
&\colhead{(12)}
&\colhead{(13)}
}
\startdata 

NGC 337 X-1&5&J005952.3$-$073447&33&  43.8 $\pm$ 6.7&32.3 $\pm$ 5.7&11.5 $\pm$ 3.5   &-0.47 $\pm$ 0.14&1.77$^{+0.35}_{-0.30}$&4.52 $\pm$ 0.69&23.2 $\pm$ 3.5&Y&ULX \\
NGC 337 X-2&6&J005951.9$-$073458&32&  14.0 $\pm$ 3.9&4.5 $\pm$ 2.2&9.5 $\pm$ 3.2  &$\ $0.36 $\pm$ 0.26&0.13$^{+0.50}_{-0.13}$&1.44 $\pm$ 0.40&7.40 $\pm$ 2.05&Y& \\
NGC 337 X-3&7&J005950.6$-$073458&19&  315.4 $\pm$ 17.8&194.8 $\pm$ 14.0&120.6 $\pm$ 11.0  &-0.24 $\pm$ 0.05&1.26$^{+0.11}_{-0.11}$&32.5 $\pm$ 1.8&167 $\pm$ 9&Y&ULX \\
NGC 337 X-4&8&J005950.4$-$073454&14&  43.2 $\pm$ 6.7&26.7 $\pm$ 5.3&16.5 $\pm$ 4.1  &-0.24 $\pm$ 0.15&1.26$^{+0.31}_{-0.29}$&4.46 $\pm$ 0.69&22.9 $\pm$ 3.6&Y&ULX \\
NGC 337 X-5&9&J005949.5$-$073523&44&  24.8 $\pm$ 5.0&18.0 $\pm$ 4.2&6.9 $\pm$ 2.6  &-0.45 $\pm$ 0.18&1.70$^{+0.45}_{-0.38}$&2.56 $\pm$ 0.52&13.1 $\pm$ 2.6&Y&ULX (c) \\
NGC 337 X-6&10&J005949.5$-$073436&10&  109.3 $\pm$ 10.5&64.7 $\pm$ 8.1&44.7 $\pm$ 6.7  &-0.18 $\pm$ 0.09&1.16$^{+0.18}_{-0.18}$&11.3 $\pm$ 1.1&57.8 $\pm$ 5.6&..&ULX \\
NGC 337 X-7&11&J005948.5$-$073457&28&  61.5 $\pm$ 7.9&41.7 $\pm$ 6.5&19.8 $\pm$ 4.5  &-0.36 $\pm$ 0.12&1.50$^{+0.27}_{-0.24}$&6.34 $\pm$ 0.81&32.5 $\pm$ 4.2&Y&ULX (b) \\
NGC 337 X-8&12&J005947.5$-$073417&45&  45.2 $\pm$ 6.8&41.7 $\pm$ 6.5&3.5 $\pm$ 2.0  &-0.85 $\pm$ 0.08&3.05$^{+0.69}_{-0.42}$&4.66 $\pm$ 0.70&23.9 $\pm$ 3.6&Y&ULX (a,c) \\
\enddata

\end{deluxetable*}

\begin{deluxetable*}{lccccccccllcc}
\setlength{\tabcolsep}{0.065in}
\addtocounter{table}{-1}
\tablewidth{0pt}
\tablecaption{Continued}
\tablehead{
\colhead{Source}
&\colhead{N}
&\colhead{\emph{Chandra} ID}
&\colhead{D}
&\colhead{Net}
&\colhead{Net}
&\colhead{Net}
&\colhead{Hardness}
&\colhead{$\Gamma$}
&\colhead{Unabs.}
&\colhead{Inferred}
&\colhead{H$\rm{\alpha}$?}
&\colhead{Notes} \\
\colhead{Name}
&\colhead{}
&\colhead{(CXOU)}
&\colhead{($''$)}
&\colhead{Counts}
&\colhead{Soft}
&\colhead{Hard}
&\colhead{Ratio}
&\colhead{}
&\colhead{Flux}
&\colhead{L$_X$}
&\colhead{}
&\colhead{} \\
\colhead{(1)}
&\colhead{(2)}
&\colhead{(3)}
&\colhead{(4)}
&\colhead{(5)} 
&\colhead{(6)}
&\colhead{(7)}
&\colhead{(8)}
&\colhead{(9)}
&\colhead{(10)}
&\colhead{(11)}
&\colhead{(12)}
&\colhead{(13)}
}
\startdata 

NGC 2805 X-9&121&J092013.1$+$640440&102& 7.7 $\pm$ 2.8&6.9 $\pm$ 2.6&0.9 $\pm$ 1.0   &-0.77 $\pm$ 0.24&2.57$^{+1.43}_{-0.75}$&0.77 $\pm$ 0.28&7.22 $\pm$ 2.65&Y&(a) \\
NGC 2805 X-10&122&J092034.5$+$640643&98& 5.5 $\pm$ 2.5&4.8 $\pm$ 2.2&0.7 $\pm$ 1.0   &-0.75 $\pm$ 0.33&2.47$^{+1.53}_{-0.92}$&0.55 $\pm$ 0.25&5.12 $\pm$ 2.30&Y&(a) \\
NGC 2805 X-11&123&J092033.2$+$640345&168& 8.6 $\pm$ 3.0&6.7 $\pm$ 2.6&1.9 $\pm$ 1.4   &-0.56 $\pm$ 0.29&1.89$^{+1.05}_{-0.63}$&0.86 $\pm$ 0.30&8.06 $\pm$ 2.81&...& \\
NGC 2805 X-12&124&J092032.1$+$640424&131& 9.7 $\pm$ 3.2&4.9 $\pm$ 2.2&4.8 $\pm$ 2.2   &-0.01 $\pm$ 0.32&0.78$^{+0.60}_{-0.60}$&0.96 $\pm$ 0.32&9.04 $\pm$ 2.96&Y&(b,c) \\
NGC 2805 X-13&125&J092020.2$+$640409&121& 4.5 $\pm$ 2.2&1.7 $\pm$ 1.4&2.8 $\pm$ 1.7   &$\ $0.24 $\pm$ 0.48&0.31$^{+0.88}_{-0.31}$&0.45 $\pm$ 0.22&4.19 $\pm$ 2.10&...& \\ 
NGC 2805 X-14&126&J092011.2$+$640723&168& 9.6 $\pm$ 3.2&8.8 $\pm$ 3.0&0.8 $\pm$ 1.0   &-0.83 $\pm$ 0.20&2.87$^{+1.13}_{-0.77}$&0.96 $\pm$ 0.32&9.00 $\pm$ 2.96&...&Star (a,b,c) \\ \\

ESO 501-23 X-1&141&J103522.2$-$244513&14& 867.0 $\pm$ 29.5&528.7 $\pm$ 23.0&338.3 $\pm$ 18.4   &-0.22 $\pm$ 0.03&1.17$^{+0.07}_{-0.06}$&84.7 $\pm$ 2.9&49.8 $\pm$ 1.7&Y&ULX    \\
\enddata
\tablecomments{
Column 1: Source name.
Column 2: Source number from Table 4.
Column 3: \chandra ID name.
Column 4: Projected distance from galactic nucleus.
Columns 5, 6, 7: Net broad, soft, and hard detected counts within 2.3'' aperture, respectively.
Column 8: Hardness Ratio, defined as (H-S)/(H+S) where H is hard counts and S is soft counts, with Gaussian errors. These are calculated using background subtracted count values, so they may not precisely map to other values in this table.
Column 9: Approximate value of the power law exponent ($\Gamma$) that would produce the observed ratio of hard to soft counts.
Column 10: Unabsorbed flux in units of 10$^{-14}$ \esc, calculated in PIMMS with a $\Gamma$=1.7 power law, including galactic absorption. Source counts were extracted from the 0.3-8 keV band, and extrapolated to 0.3-10 keV.
Column 11: Inferred luminosity of the source, at the galactic distance, in units of 10$^{38}$ \es. This is only valid for sources that are in fact associated with the galaxy.
Column 12: Does this object coincide with H$\alpha$ emission from MDM imaging? 
Column 13: ULX indicates a ULX candidate. (a) indicates a match in the UNSO-B1.0 catalog, (b) indicates a match in the WISE point source catalog, and (c) indicates a match in the SDSS DR8 catalog.
}
\end{deluxetable*}

\clearpage

\end{landscape}

\clearpage

\end{document}